\documentclass[pra,twocolumn,superscriptaddress,showpacs,amsmath,amstex,amssymb,citeautoscript]{revtex4-2}

\usepackage[T1]{fontenc}
\usepackage[utf8]{inputenc}
\usepackage{lipsum}
\usepackage{amsmath}
\usepackage{amssymb}
\usepackage{bbm}
\usepackage{braket}
\usepackage{xcolor}
\usepackage{pifont}
\usepackage[mathscr]{euscript}
\usepackage[shortlabels]{enumitem}
\usepackage[caption=false]{subfig}
\usepackage{overpic}

\usepackage{tikz-feynman}
\usepackage{tikz-cd}

\usepackage{graphicx}
\usepackage{lipsum}
\allowdisplaybreaks
\usepackage{float}
\usepackage{graphicx}
\usepackage{dsfont}
\usepackage{comment}
\usepackage[colorlinks=true]{hyperref}  
\hypersetup{
    bookmarks=true,         
    unicode=false,          
    pdftoolbar=true,        
    pdfmenubar=true,        
    pdffitwindow=false,     
    pdfstartview={FitH},    
    pdftitle={Vestigial phases of magnetism and triplet superconductivity},    
    pdfauthor={Verghis et al.},     
    pdfsubject={},   
    pdfcreator={},   
    pdfproducer={}, 
    pdfkeywords={} {} {}, 
    pdfnewwindow=true,      
    colorlinks=true,       
    linkcolor=blue, 
    citecolor=blue,        
    filecolor=magenta,      
    urlcolor=blue           
}

\newcommand{\equref}[1]{Eq.~(\ref{#1})}
\newcommand{\equsref}[2]{Eqs.~(\ref{#1}) and (\ref{#2})}
\newcommand{\secref}[1]{Sec.~\ref{#1}}
\newcommand{\figref}[1]{Fig.~\ref{#1}}
\newcommand{\refcite}[1]{Ref.~\onlinecite{#1}}

\newcommand{\tableref}[1]{Table~\ref{#1}}

\newcommand{\diff}{\mathrm{d}}
\newcommand{\sign}{\,\text{sign}}
\renewcommand{\approx}{\simeq}
\renewcommand{\Re}{\text{Re}}
\renewcommand{\Im}{\text{Im}}
\newcommand{\D}{\Delta_{\pmb{k}}}

\renewcommand{\vec}[1]{\boldsymbol{#1}}
\newcommand{\Tr}{\text{Tr}}

\definecolor{wrongultramarine}{rgb}{1,0.5,0}

\definecolor{red_lines}{rgb}{0.65, 0, 0}

\renewcommand{\pmb}[1]{\boldsymbol{#1}}

\begin{document}

\title{Vestigial pairing from fluctuating magnetism and triplet superconductivity}

\author{Yanek Verghis}
\affiliation{Institute for Theoretical Physics III, University of Stuttgart, 70550 Stuttgart, Germany}

\author{Denis Sedov}
\affiliation{Institute for Theoretical Physics III, University of Stuttgart, 70550 Stuttgart, Germany}

\author{Jakob We\ss ling}
\affiliation{Institute for Theoretical Physics, University of Innsbruck, Innsbruck A-6020, Austria}

\author{Prathyush P. Poduval}
\affiliation{Donald Bren School of Information and Computer Science, University of California, Irvine, CA 92697, USA}

\author{Mathias S.~Scheurer}
\affiliation{Institute for Theoretical Physics III, University of Stuttgart, 70550 Stuttgart, Germany}

\begin{abstract}
We study the finite-temperature vestigial superconducting phases of a two-dimensional system of fluctuating spin-triplet pairing and spin magnetism. Denoting the respective primary order parameters by $\vec{d}$ and $\vec{N}$, which are not long-range ordered at finite temperature, the composite fields $\phi_{dd} = \vec{d}\cdot\vec{d}$ and $\phi_{dN} = \vec{d}\cdot\vec{N}$ are spin-rotation invariant and can condense at finite temperature. Using a large-$N$ approach that respects the Mermin-Wagner theorem, we here derive the phase diagram which features two vestigial superconductors: $(A)$ a charge-$4e$ superconductor with $\phi_{dd}\neq 0$ and $\phi_{dN} =0$ and $(B)$ a charge-$2e$ state with $\phi_{dN} ,\phi_{dd}\neq 0$. We analyze the temperature and coupling-constant dependent properties of these two superconductors using a perturbative approach and a variational Hartree-Fock study. This reveals non-trivial spectra in the superconductors, which result from the fundamental building blocks being distinct from the usual Cooper pairs---in phase $(A)$, the elementary bosons are bound states of four electrons and, in phase $(B)$, of three electrons and a hole. This work complements the previous study [\href{https://www.nature.com/articles/s41467-024-45950-4}{Nat.~Commun.~\textbf{15}, 1713 (2024)}], which focused on the properties of phase $(B)$.
\end{abstract}

\maketitle
\section{Introduction}
Vestigial order \cite{nieQuenchedDisorderVestigial2014,fernandesIntertwinedVestigialOrder2019} has become an important theoretical concept for our understanding of the multitude of phases appearing in the phase diagrams of many strongly correlated quantum materials. As opposed to being solely the result of the competition between different ordering tendencies, the idea is that at least certain aspects of complex diagrams are better thought of as ``intertwined'' \cite{fradkinColloquiumTheoryIntertwined2015} order parameters with a common, more fundamental origin. To illustrate the concept, let us denote by $\eta_j$ a set of (real or complex) primary order-parameter components, which when condensed, $\braket{\eta_j} \neq 0$, would break a certain set of symmetries. However, even when $\braket{\eta_j} = 0$, it is possible that composite order parameters, e.g., bilinears $\phi_{j,j’} = \eta_j \eta_{j’}$ or higher powers, are still long-range or quasi-long-range ordered; this defines a ``vestigial phase’’ where only a subset of the symmetries are broken compared to the phase where the involved $\eta_j$ are ordered individually. 

For instance, \refcite{PhysRevB.85.024534} discussed the situation, relevant to the iron pnictides, where $(\eta_{j})_{j=1,2,3} = \vec{M}_x$ and $(\eta_{j})_{j=4,5,6} = \vec{M}_y$ are the magnetic order parameters with ordering vectors $\vec{Q} = (\pi,0)^T$ and $\vec{Q} = (0,\pi)^T$, respectively. This allows to define the Ising-nematic composite order parameter $\phi_I = \vec{M}_x^2 - \vec{M}^2_y$ and the vestigial phase where $\braket{\vec{M}_{x,y}} =0$ while $\braket{\phi_I} \neq 0$, i.e., a state with preserved spin-rotation symmetry but characterized by a spontaneous distortion from tetragonal to orthorhombic. Another example \cite{bergCharge4eSuperconductivityPairdensitywave2009} is the charge-$4e$ vestigial superconductor that arises once a finite-momentum superconducting order parameter $\Delta_{\vec{Q}}$ with charge $2e$ is ``melted'', while the product $\phi = \Delta_{\vec{Q}} \Delta_{-\vec{Q}}$ remains condensed. In fact, charge-$4e$ pairing of this and other forms has attracted a lot of attention recently \cite{grinenkoStateSpontaneouslyBroken2021,geDiscoveryCharge4eCharge6e2022,fernandesChargeSuperconductivityMulticomponent2021,jianChargeSuperconductivityNematic2021,zengPhasefluctuationInducedTimeReversal2021,songPhaseCoherencePairs2022,PhysRevB.107.064501,chungBerezinskiiKosterlitzThoulessTransitionTransport2022,jiangCharge4eSuperconductors2017,LI20242328,gnezdilovSolvableModelCharge2022,garaudEffectiveModelMagnetic2022,Sixfermions,PhysRevB.108.054517,zhouChernFermiPocket2022,curtisStabilizingFluctuatingSpintriplet2022,Main_paper,2024npjQM...9...66W,2024PhRvL.133m6002L,PhysRevB.110.054519,PhysRevB.109.134514,PhysRevB.107.224503,PhysRevB.109.144504}.
 
As it is most relevant to us here, lets focus in more detail on the scenario proposed in \refcite{Main_paper}, where an interesting set of vestigial phases emerges from primary order parameters given by the three (real) components of the spin magnetization $\vec{N}$ and the three (complex) components $\vec{d}$ of a superconducting triplet order parameter. This is motivated by the phenomenology of correlated van der Waals multi-layer systems, where there is experimental evidence for the presence of superconductivity and magnetism in the same density range \cite{wongCascadeElectronicTransitions2020, zondinerCascadePhaseTransitions2020,parkTunableStronglyCoupled2021,haoElectricFieldTunable2021,kimEvidenceUnconventionalSuperconductivity2022,ohEvidenceUnconventionalSuperconductivity2021,morissetteElectronSpinResonance2022} and even for microscopic coexistence \cite{linZerofieldSuperconductingDiode2022,rtg_chiral_sc_experiment,MoTe2_unconv_exp}; this, together with \cite{PabloMagneticField,sainz-cruzJunctionsSuperconductingSymmetry2022,ohEvidenceUnconventionalSuperconductivity2021}, make pairing in the triplet channel very natural. Reference \onlinecite{Main_paper} focused on one particular possible vestigial superconducting state and showed that, although the primary triplet pairing order would be fully gapped, its vestige can naturally realize a crossover from $V$-shaped to $U$-shaped in the density of states (DOS), similar to what is seen experimentally \cite{kimEvidenceUnconventionalSuperconductivity2022, ohEvidenceUnconventionalSuperconductivity2021}.

In this work, we complement the analysis of \refcite{Main_paper} in multiple ways. First, by studying the vestigial phase diagram. To control the calculation, we use a large-$N$ theory where the number of components of $\vec{d}$ and $\vec{N}$ is generalized from $3$ to $N$ and demonstrate that, depending on parameters, indeed both naturally expected vestigial superconducting phases can be realized: Defining the spin-rotation invariant composite order parameters $\phi_{dd} = \vec{d}\cdot\vec{d}$ and $\phi_{dN} = \vec{d}\cdot\vec{N}$, phase $(A)$ has $\braket{\phi_{dd}} \neq 0$ while $\braket{\phi_{dN}} = 0$ and, thus, constitutes a charge-$4e$ superconductor; in contrast, in phase $(B)$, we have $\braket{\phi_{dN}} \neq 0$ and, thus, also $\braket{\phi_{dd}} \neq 0$, defining a vestigial charge-$2e$ state. We analyze the order of the thermal phase transitions as well as of the transitions between the different phases driven by variation of non-thermal parameters. 

Second, we also study the spectral properties of the two vestigial states with a particular focus on phase $(A)$, which was not considered in \refcite{Main_paper}, and contrast it with phase $(B)$. To this end, we use two complementary approaches which yield overall consistent results: leading-order perturbation theory, keeping all dynamical correlations and life-time broadening effects, and a self-consistent Hartree-Fock approach, which can capture non-perturbative effects but neglects frequency dependencies. We discuss in detail the resulting self-consistency equations, which have a qualitatively different form than those known from the standard theory of (non-vestigial) superconductivity, and extend the analysis of phase $(B)$ in \refcite{Main_paper} to finite momentum transfer $\vec{q}$. Both superconductors display a partially suppressed DOS which, upon lowering temperature and increasing the coupling constants, eventually becomes fully gapped. On top of this, depending on parameters, phase $(A)$ exhibits additional subgap peaks.
We note that the possibility of vestigial $\vec{d}\cdot\vec{d}$ pairing was mentioned before \cite{curtisStabilizingFluctuatingSpintriplet2022}, but, to the best of our knowledge, no analysis of its electronic spectrum as we present here was provided. 

The remainder of the manuscript is organized as follows. In \secref{BosonicTheory}, we derive the phase diagram of vestigial orders. The electronic spectral properties of all phases are then addressed in \secref{EffElectrTheory}, starting with a perturbative study of the self energies (\secref{SelfEnergy}) before using a Hartree-Fock approach (\secref{HartreeFockTheory}). Finally, our findings are summarized in \secref{Summary}.

\section{Bosonic theory}\label{BosonicTheory}
As already explained above, we here consider a two-dimensional (2D) system with strong tendencies towards both triplet superconductivity and spin magnetism. The associated order parameter fields $\pmb{d}_q$ and $\pmb{N}_q$ couple to the fermions as
\begin{align}\begin{split}
        S_c =\lambda \int_{k, q} \bigl[&\bar{c}_{k-q} \pmb{s} \cdot\pmb{N}_{q}\tau_z c_{k} \\ &+\left(\bar{c}_{k-q} \pmb{s}\cdot \pmb{d}_{q} is_{y} \tau_{y} \bar{c}_{-k}+\text{H.c.}\right)\bigr].
        \end{split}
\end{align}
Here the electronic field operators of spin $s=\uparrow,\downarrow$ (Pauli matrices $\pmb{s}$) and valley $\tau=\pm$ are $c_{k,s,\tau}$; $k=(i\omega_n,\pmb{k})$ comprises Matsubara frequencies and 2D momentum. We include a valley quantum number since, similar to \refcite{Main_paper}, the work is motivated by the superconducting physics of correlated van der Waals stacks: in these systems, there are multiple indications for the simultaneous emergence of the aforementioned ordering tendencies \cite{wongCascadeElectronicTransitions2020, zondinerCascadePhaseTransitions2020,parkTunableStronglyCoupled2021,haoElectricFieldTunable2021,kimEvidenceUnconventionalSuperconductivity2022,ohEvidenceUnconventionalSuperconductivity2021,morissetteElectronSpinResonance2022,linZerofieldSuperconductingDiode2022,scammellTheoryZerofieldSuperconducting2022,PabloMagneticField,sainz-cruzJunctionsSuperconductingSymmetry2022,ohEvidenceUnconventionalSuperconductivity2021,rtg_chiral_sc_experiment,MoTe2_unconv_exp}. However, we note that our analysis can be straightforwardly generalized to other systems without valleys. The main difference will be that the superconducting triplet order parameter then necessarily depends on the relative momentum $\vec{k}$ of the paired fermions.

Being interested in finite-temperature physics, where the primary order parameter fields $\pmb{d}_q, \pmb{N}_q$ fluctuate, we treat them as dynamical bosons, akin to the well-known spin-fermion model \cite{SFM}. Their bare dynamics is governed by
\begin{equation}
    S_b=\int_q  \left[\chi_{N}^{-1}(q) \pmb{N}_{q} \cdot\pmb{N}_{-q}+\chi_{d}^{-1}(q) \pmb{\bar{d}}_{q}\cdot \pmb{d}_{q}\right],  \label{BareBosonicAction}
\end{equation}
where the susceptibilities are $\chi_{\mu}=\bar{\chi}_{\mu}/(r_\mu+\Omega_n^2+v_\mu^2\pmb{q}^2),\mu=d,N$, and $q=(i\Omega_n,\pmb{q})$ comprises the bosonic Matsubara frequencies and momentum. We will set $\bar{\chi}_{\mu}=1$ in the following by rescaling the fields (and $\lambda$). There are additional terms of higher order in the fields, which we will come back to shortly. 

Integrating out the electrons, described by the bare action $S_e=\int_{k}\bar{c}_{k,s,\tau}\left(-i\omega_n+\epsilon_{k,\tau}\right)c_{k,s,\tau}$, one can derive an effective theory $S^{\text{eff}}_B$ for the bosons only. As usual, the form of this theory can be derived based on symmetries and further constrained by locality. Keeping terms only up to quartic order in fields or gradients, we conclude from the symmetries listed in \tableref{SymmetriesAndReps} that $S^{\text{eff}}_B = S_b + S_V$ with $S_V=\int_xV\left(\pmb{d}(x),\pmb{N}(x) \right)$ where   
\begin{equation}
    V=b_1(\pmb{\bar{d}}\pmb{d})^2-b_2|\pmb{d}\pmb{d}|^2+b_3\pmb{N}^4-c_1|\pmb{d}\pmb{N}|^2+c_2\left(\pmb{\bar{d}}\pmb{d}\right)\pmb{N}^2. \label{Potential}
\end{equation}
Note that we have absorbed the renormalization of the susceptibilities $\chi_\mu$, caused by integrating out the electrons, into a redefinition of their parameters ($\bar{\chi}_\mu$, $r_\mu$, and $v_\mu$) and refrained from changing the symbol to keep the notation compact.

\begin{table}[tb]
\begin{center}
\caption{Summary of symmetries $g$ and how they act on the important fields in our theory. Note that all symmetries are linear, safe for time-reversal $\Theta$ which is anti-linear; $R_{\vec{\varphi}}$ is the $3\times 3$ rotation matrix obeying $e^{-i\vec{\varphi}\cdot\vec{s}} \vec{s} e^{i\vec{\varphi}\cdot\vec{s}} = R(\vec{\varphi}) \vec{s}$.}
\label{SymmetriesAndReps}\begin{ruledtabular}\begin{tabular}{cccccc} 
$g$ & $c_{\vec{k}}$ & $\vec{N}$ & $\vec{d}$ & $\phi_{dd}$ & $\phi_{dN}$  \\ \hline
$U(1)$ & $e^{i\varphi}c_{\vec{k}}$ & $\vec{N}$ & $e^{-2i\varphi}\vec{d}$ & $e^{-4i\varphi}\phi_{dd}$ & $e^{-2i\varphi}\phi_{dN}$  \\
SO(3) & $e^{i\vec{\varphi}\cdot\vec{s}}c_{\vec{k}}$  & $R_{\vec{\varphi}} \vec{N}$ & $R_{\vec{\varphi}} \vec{d}$ & $\phi_{dd}$ & $\phi_{dN}$  \\
$C_{2z}$ & $\tau_x c_{-\vec{k}}$  & $-\vec{N}$ & $-\vec{d}$ & $\phi_{dd}$ & $\phi_{dN}$   \\
$\Theta$ & $i s_y \tau_x c_{-\vec{k}}$  & $\vec{N}$ & $-\vec{d}^*$ & $\phi^*_{dd}$ & $-\phi^*_{dN}$ 
\end{tabular}\end{ruledtabular}
\end{center}
\end{table}

\subsection{Large-$N$ approach}\label{LargeN}
Our goal here is to study the resulting vestigial phases from the effective bosonic action $S^{\text{eff}}_B$. To this end, we will follow \refcite{PhysRevB.85.024534} and decouple the quartic terms in \equref{Potential} using two real Hubbard-Stratonovich fields---$\psi_N$ for $\vec{N}^2$ and $\psi_d$ for $\pmb{\bar{d}}\vec{d}$---as well as two complex fields, $\phi_{dd}$ and $\phi_{dN}$, for $\vec{d}\vec{d}$ and $\vec{d}\vec{N}$, respectively. With this, the action becomes
\begin{align}\begin{split}
    S^{\text{HS}}_B =  S_b + \int_x &\Bigl[ \frac{b_3 \psi_d^2 + b_1 \psi_N^2 + c_2 \psi_d \psi_N}{4b_1b_3 - c_2^2} + \frac{|\phi_{dd}|^2}{b_2} \\ &\quad + \frac{|\phi_{dN}|^2}{c_1} + i \psi_d \pmb{\bar{d}}\pmb{d} + i \psi_N \pmb{N}^2  \\
    &\quad + ( \phi_{dd} \vec{d}\vec{d} + \phi_{dN} \vec{d}\vec{N} + \text{H.c.}) \Bigr]. \label{SBHS}
\end{split}\end{align}
Here we assume positive $b_2,c_1$, but note that the final saddle-point equations [see \equref{SaddlePointEquations} below] we will end up with, will hold for any combination of their signs. Since $c_2 \neq 0$ does not lead to any additional phases, we will set $c_2=0$ in the following for simplicity. Integrating out $\vec{d}$ and $\vec{N}$ in \equref{SBHS}, we arrive at the effective action of the Hubbard-Stratonovich fields,
\begin{align}\begin{split}
    S^{\text{HS}}_{\psi\phi} &=  \int_q \Bigl\{ \frac{\psi_d^2}{4 b_1} + \frac{\psi_N^2}{4 b_3} + \frac{|\phi_{dd}|^2}{b_2} + \frac{|\phi_{dN}|^2}{c_1} \\ & +  \frac{3}{2} \ln \bigl[ (\chi_N^{-1} + i \psi_d) \bigl[(\chi_d^{-1} + i \psi_d) - 4 |\phi_{dd}|^2 \bigr] \\
    &- (\chi_d^{-1} + i \psi_d) |\phi_{dN}|^2 + \bar{\phi}_{dd} \phi_{dN}^2 + \phi_{dd} \bar{\phi}_{dN}^2 \bigr] \Bigr\}, \label{FinalHSAction}
\end{split}\end{align}
where we switched to a frequency-momentum representation (but suppress the momentum arguments of the Hubbard-Stratonovich fields for notational brevity).

To proceed, we will employ a large-$N$ approach, where the three-component fields $\vec{d}$ and $\vec{N}$ are promoted to $N$ component fields, while keeping full $O(N)$ invariance in $S^{\text{eff}}_B$. To treat the quadratic and quartic terms on equal footing, we will further rescale the coupling constants in \equref{Potential} as $b_j \rightarrow b_j/N$ and $c_j \rightarrow c_j/N$. Overall, this will just yield a global prefactor of $N$ in \equref{FinalHSAction} such that the saddle point approximation with respect to $\psi_d$, $\psi_N$, $\phi_{dd}$, $\bar{\phi}_{dd}$, $\phi_{dN}$, and $\bar{\phi}_{dN}$ becomes exact in the limit $N\rightarrow \infty$.

Out of the six associated saddle-point equations, only four are independent and read as
\begin{subequations}\begin{align}
    \frac{\psi_d^0}{b_1} &= \int_q \frac{-2i (\chi_d^{-1} + i \psi^0_d)(\chi_N^{-1} + i \psi_N) + i |\phi^0_{dN}|^2 }{Q[\chi_d,\chi_N,\psi^0_d,\psi^0_N,\phi^0_{dd},\phi^0_{dN}]}, \label{FirstSPE} \\
    \frac{\psi^0_N}{b_3} &= \int_q \frac{-i (\chi_d^{-1} + i \psi^0_d)^2 + 4i |\phi^0_{dd}|^2 }{Q[\chi_d,\chi_N,\psi^0_d,\psi^0_N,\phi^0_{dd},\phi^0_{dN}]}, \label{SecondSPE} \\
    \frac{\phi_{dd}^0}{b_2} &= \int_q \frac{2 \phi_{dd}^0 (\chi_N^{-1} + i \psi^0_N) - (\phi^0_{dN})^2/2 }{Q[\chi_d,\chi_N,\psi^0_d,\psi^0_N,\phi^0_{dd},\phi^0_{dN}]}, \label{ThirdSPE} \\
    \frac{\phi_{dN}^0}{c_1} &= \int_q \frac{\phi_{dN}^{0} (\chi_d^{-1} + i \psi^0_d)/2 -  \phi^0_{dd}\bar{\phi}^0_{dN}}{Q[\chi_d,\chi_N,\psi^0_d,\psi^0_N,\phi^0_{dd},\phi^0_{dN}]},
\end{align}\label{SaddlePointEquations}\end{subequations}
where the denominator is identical in all cases and given by
\begin{align}\begin{split}
    Q &= (\chi_N^{-1} + i \psi^0_N) \left[ (\chi_d^{-1} + i \psi^0_d)^2 - 4 |\phi^0_{dd}|^2\right] \\ & \, - (\chi_d^{-1} + i \psi^0_d) |\phi^0_{dN}|^2 + \bar{\phi}^0_{dd} (\phi^0_{dN})^2 + \phi^0_{dd} (\bar{\phi}_{dN}^0)^2.
\end{split}\end{align}
First note that the solutions of $\psi_d^0,\psi_N^0$ are purely imaginary and their impact is to renormalize the bosonic masses,
\begin{equation}
    r_\mu \, \rightarrow \, \tilde{r}_\mu = r_\mu + i \psi^0_\mu. \label{MassRenormalization}
\end{equation}
As $\phi_{dN}$ transforms non-trivially under U(1) gauge transformations [cf.~\tableref{SymmetriesAndReps}], we can choose a gauge where $\phi^0_{dN} \in \mathbb{R}$, without loss of generality. From \equref{ThirdSPE} we can then immediately see that this also implies $\phi^0_{dd} \in \mathbb{R}$. As such, we will take both $\phi^0_{dN}$ and $\phi^0_{dd}$ to be real in the following. Another important observation is that $\phi^0_{dd} = 0$ generically implies $\phi^0_{dN} = 0$, which also directly follows from \equref{ThirdSPE}. This is natural, since $\braket{\vec{d}\cdot \vec{N}} \neq 0$, which defines a charge-$2e$ superconductor \cite{Main_paper}, is also expected to lead to $\braket{\vec{d}\cdot \vec{d}} \neq 0$ without fine-tuning. Put differently, if we already have $\phi^0_{dN} \neq 0$, tuning on and off $\phi^0_{dd}$ does not change any symmetries.

We emphasize that our large-$N$ theory respects the Mermin-Wagner theorem and we thus expect to always find $\braket{\vec{d}} = \braket{\vec{N}} =0$ at finite temperatures. To check this, one can extract the effective bosonic susceptibility $\underline{\chi}$ from the saddle-point limit of \equref{SBHS}, which reads as $S^{\text{HS},0}_B  = \int_q \bar{\vec{B}}_q \underline{\chi}(q) \vec{B}_q$, where $\vec{B}_q = (\vec{N}_q,\vec{d}_{q},\pmb{\bar{d}}^T_{-q})^T$, $\bar{\vec{B}}_q = (\vec{N}_{-q},\pmb{\bar{d}}_{q},\vec{d}_{-q}^T)$ and
\begin{equation}
    \underline{\chi}(q) = \begin{pmatrix}
        \widetilde{\chi}^{-1}_{N}(q) & \phi^0_{dN}/2 &  \bar{\phi}^0_{dN}/2 \\
        \bar{\phi}^0_{dN}/2 & \widetilde{\chi}^{-1}_{d}(q)/2 & \bar{\phi}^0_{dd} \\
        \phi^0_{dN}/2 & \phi_{dd}^0 & \widetilde{\chi}^{-1}_{d}(q)/2 
    \end{pmatrix}. \label{BosonicSusceptibility}
\end{equation}
Here, $\widetilde{\chi}^{-1}_{\mu} = \chi^{-1}_{\mu} + i \psi^0_\mu$ are the renormalized inverse susceptibilities as per \equref{MassRenormalization}. We have checked in our explicit numerical calculations below that $\underline{\chi}(q)$ is always positive definite at the finite-temperature saddle points. This implies that the primary order parameters $\braket{\vec{d}}$ and $\braket{\vec{N}}$ are always uncondensed. Mathematically, this is related to the fact that $Q = 4 \det \underline{\chi}$ such that the integrals on the right-hand sides of \equref{SaddlePointEquations} would become divergent (due to some $q\neq \infty$) if one of the eigenvalues of $\underline{\chi}$ changed sign.

\begin{figure*}[tb]
    \centering
    \includegraphics[width=\linewidth]{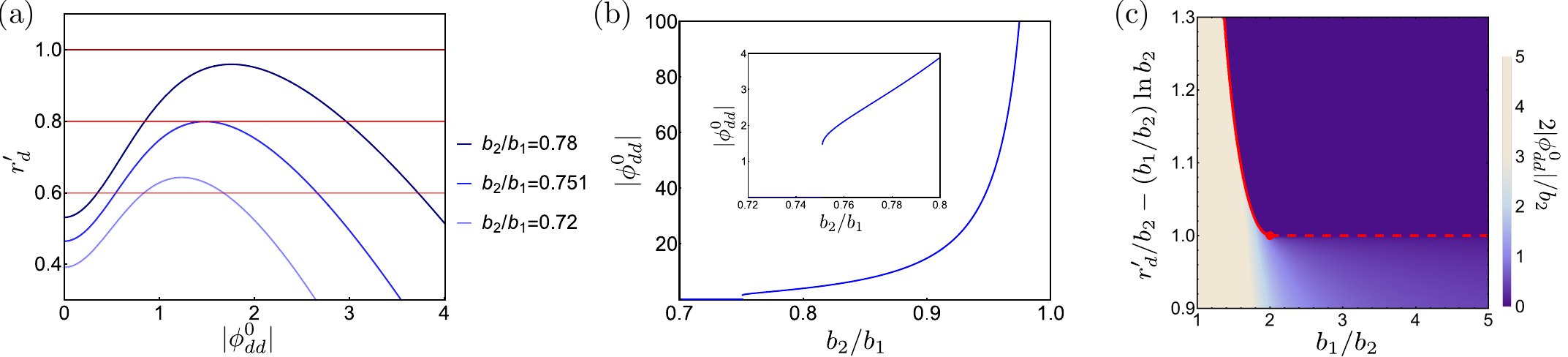}
    \caption{(a) Visual representation of \equref{SEToSolve} with red (blue) referring to left-hand (right-hand) side. (b) Solution of \equref{SEToSolve} with lower free energy as a function of $b_2/b_1$ for $r_d'=1$ and $b_1=1$. For $b_2>b_1$, the theory is unstable. The inset is a zoom-in on the transition region. (c) Phase diagram for $\phi^0_{dN} = 0$, where solid (dashed) red lines refer to first (second) order phase transitions. Here, we have rescaled all parameters in \equref{SEToSolve} so that the phase diagram depends only on the single parameter $b_1/b_2$. As explained in the text, $r_d'$ can be thought of as (being in a monotonic relation with) temperature within our theory.}
    \label{fig:SpecialCase}
\end{figure*}

Taken together, we already see that there are (at most) three phases at finite temperature: 
\begin{enumerate}
    \item[(A):]here only $\phi^0_{dd}$ is non-zero, while $\phi^0_{dN} = 0$. This defines a symmetric, charge-$4e$ superconductor, which can intuitively be thought of as the result of two Cooper pairs, each in a spin-triplet state, to form an overall spin-zero configuration.
    \item[(B):]this state is characterized by $\phi^0_{dN},\phi^0_{dd} \neq 0$. Although preserving all symmetries and just being a charge-$2e$ superconductor, it differs from the BCS state as it should be thought of as the result of pairing three particles and a hole. The analysis of \refcite{Main_paper} focused on this state.
    \item[(C):]here, both saddle-point values vanish, $\phi^0_{dN} = \phi^0_{dd} = 0$, and we do not obtain a superconductor. As mentioned in \refcite{Main_paper}, one can still have vestigial broken time-reversal symmetry; however, we will not consider this aspect here.
\end{enumerate}

\subsection{Solution and phase diagram}\label{PhaseDiagram}
We next proceed to explicitly solve the saddle-point equations (\ref{SaddlePointEquations}), focusing on the classical, high-temperature limit where only the zeroth Matsubara frequency needs to be taken into account. Specifically, we replace $\int_q \rightarrow \int \diff^2 \vec{q} /(2\pi)^2$. As \equsref{FirstSPE}{SecondSPE} are both logarithmically divergent in the ultra-violet, we introduce a momentum cutoff $\Lambda$ for both of them. Rewriting them as equations for $\tilde{r}_\mu$, we can just absorb a $\Lambda$-dependent, divergent term via 
\begin{equation}
    r_d \rightarrow r'_d = r_d + \frac{b_1}{(2\pi)^2} \int  \frac{2}{q^2+q_0^2} q \diff q \label{AbsorbDivergentTerm}
\end{equation}
and similarly for $r_N$. Then we can set $\Lambda \rightarrow \infty$ in the remaining, convergent integral. In this way, we have absorbed all non-universal constants into $r'_\mu$, which are our remaining, thermal control parameters, driving the vestigial phase transitions. The additional parameter $q_0$ introduced in \equref{AbsorbDivergentTerm} is used as a momentum scale by writing $q = q_0 p$, where $p$ is dimensionless. In the following, we further rescale 
\begin{align*}
    &\hat{\tilde{r}}_\mu = \frac{\tilde{r}_\mu}{(v_\mu q_0)^2}, \quad \hat{\phi}^0_{dN} = \frac{\phi_{dN}^0}{v_Nq_0 v_d q_0}, \quad \hat{\phi}^0_{dd} = \frac{\phi_{dd}^0}{(v_d q_0)^2}, \\
    &\hat{b}_1 = \frac{b_1}{2\pi v_d^4 q_0^2}, \hat{b}_2 = \frac{b_2}{2\pi v_d^4 q_0^2},     \hat{b}_3 = \frac{b_3}{2\pi v_N^4 q_0^2},  \hat{c}_1 = \frac{c_1}{2\pi v_N^2 v_d^2 q_0^2},
\end{align*}
to write the resulting saddle-point equations in compact form, 
\begin{widetext}
\begin{subequations}\begin{align}
    \tilde{r}_d &= r_d'+ b_1 \int_0^\infty  \left[\frac{2(\tilde{r}_N + p^2)(\tilde{r}_d + p^2) - (\phi_{dN}^0)^2}{P(p,\tilde{r}_d,\tilde{r}_N,\phi^0_{dd},\phi^0_{dN})} -  \frac{2}{p^2+1}\right] p \, \diff p, \\
    \tilde{r}_N &= r_N'+ b_3 \int_0^\infty \left[\frac{(\tilde{r}_d + p^2)^2 - 4(\phi_{dd}^0)^2}{P(p,\tilde{r}_d,\tilde{r}_N,\phi^0_{dd},\phi^0_{dN})} -  \frac{1}{p^2+1}\right] p \, \diff p ,  \\
    \phi^0_{dd} & = b_2  \int_0^\infty  \frac{2 \phi_{dd}^0 (\tilde{r}_N + p^2) - (\phi_{dN}^0)^2/2}{P(p,\tilde{r}_d,\tilde{r}_N,\phi^0_{dd},\phi^0_{dN})} p \, \diff p, \\
    \phi^0_{dN} & = c_1  \int_0^\infty  \frac{2 \phi_{dN}^0 (\tilde{r}_d + p^2)/2 - \phi_{dd}^0 \phi_{dN}^0}{P(p,\tilde{r}_d,\tilde{r}_N,\phi^0_{dd},\phi^0_{dN})} p \, \diff p, \label{SCEd}
\end{align}\label{SCE}\end{subequations}
where we introduced the dimensionless form of the denominator
\begin{equation}
    P(p,\tilde{r}_d,\tilde{r}_N,\phi^0_{dd},\phi^0_{dN}) = (\tilde{r}_N + p^2) \left[ (\tilde{r}_d + p^2)^2 - 4 (\phi_{dd}^0)^2 \right] - (\tilde{r}_d + p^2) (\phi^0_{dN})^2 + 2 \phi_{dd}^0 (\phi_{dN}^0)^2
\end{equation}
and we dropped the additional hats to keep the notation compact.
\end{widetext}
Let us start our analysis with phase $(A)$, where $\phi^0_{dN} = 0$ which automatically solves \equref{SCEd}. Meanwhile, the remaining three saddle-point equations simplify considerably, allowing for a straightforward evaluation of the momentum integrals. We find
\begin{subequations}
\begin{align}
    \tilde{r}_d &= r_d'- \frac{b_1}{2} \ln (\tilde{r}_d^2 - 4(\phi_{dd}^0)^2), \label{SimplifiedSEa} \\
    \tilde{r}_N &= r_N' - \frac{b_3}{2}  \ln \tilde{r}_N, \\
    \phi^0_{dd} & = \phi^0_{dd} \frac{b_2}{2|\phi_{dd}^0|} \text{arcoth}\left(\frac{\tilde{r}_d}{2|\phi_{dd}^0|}\right).  \label{SimplifiedSEd}
\end{align}\label{SimplifiedSE}\end{subequations}
Since the second equation is decoupled from the others, we can ignore it in the following. We either have $\phi^0_{dd} = 0$, which trivially solves \equref{SimplifiedSEd} and $\tilde{r}_d(r_d')$ can be determined from \equref{SimplifiedSEa}, or $\phi^0_{dd} \neq 0$. Note that \equref{SimplifiedSEd} immediately implies $|\phi_{dd}^0| < \tilde{r}_d/2$, which, as already seen on a more general level above, guarantees that $\underline{\chi}$ in \equref{BosonicSusceptibility} is positive definite, in line with the Mermin-Wagner theorem. If $\phi^0_{dd} \neq 0$, we can cancel it out on both sides of \equref{SimplifiedSEd}, solve this equation for $\tilde{r}_d$, and plug it back into \equref{SimplifiedSEa} yielding
\begin{equation}
    r_d' = 2|\phi_{dd}^0| \coth \left( \frac{2|\phi_{dd}^0|}{b_2} \right) + b_1 \ln \left( \frac{2 |\phi_{dd}^0|}{\sinh (2 |\phi_{dd}^0|/b_2)} \right). \label{SEToSolve}
\end{equation}
In \figref{fig:SpecialCase}(a), we illustrate solving \equref{SEToSolve} graphically. Recalling that $r_d'$ encodes the temperature $T$ in our theory, we can see that for large $r_d'$ (high $T$) and small $b_2/b_1$, there is no intersection such that we are left with the solution $\phi^0_{dd} = 0$ of \equref{SimplifiedSEd}. If $T$ is lowered (smaller $r_d'$) or $b_2/b_1$ increased, the two graphs do intersect. For the set of parameters used, the intersection happens at a finite value of $\phi^0_{dd}$, pointing at a first-order transition into the charge-$4e$ superconductor of phase $(A)$. When there are two intersections in \figref{fig:SpecialCase}(a), the one with larger $\phi^0_{dd}$ has lower energy among the two, which we have checked by computing the free energy and is also required by stability. The evolution of this solution of \equref{SEToSolve} with $b_2/b_1$ is shown in \figref{fig:SpecialCase}(b). The divergence as $b_2/b_1 \rightarrow 1$ is related to the fact that the theory becomes unstable.

In \figref{fig:SpecialCase}(c), we show the resulting phase diagram, where we took into account the free energy of the different possible solutions of \equref{SimplifiedSE} and only plot the stable configuration. As expected, the superconducting phase appears at lower temperatures. We do indeed get a first order transition for small $b_1/b_2$ while it becomes second order for $b_1 > 2b_2$.

\begin{figure}[tb]
    \centering
    \includegraphics[width=\linewidth]{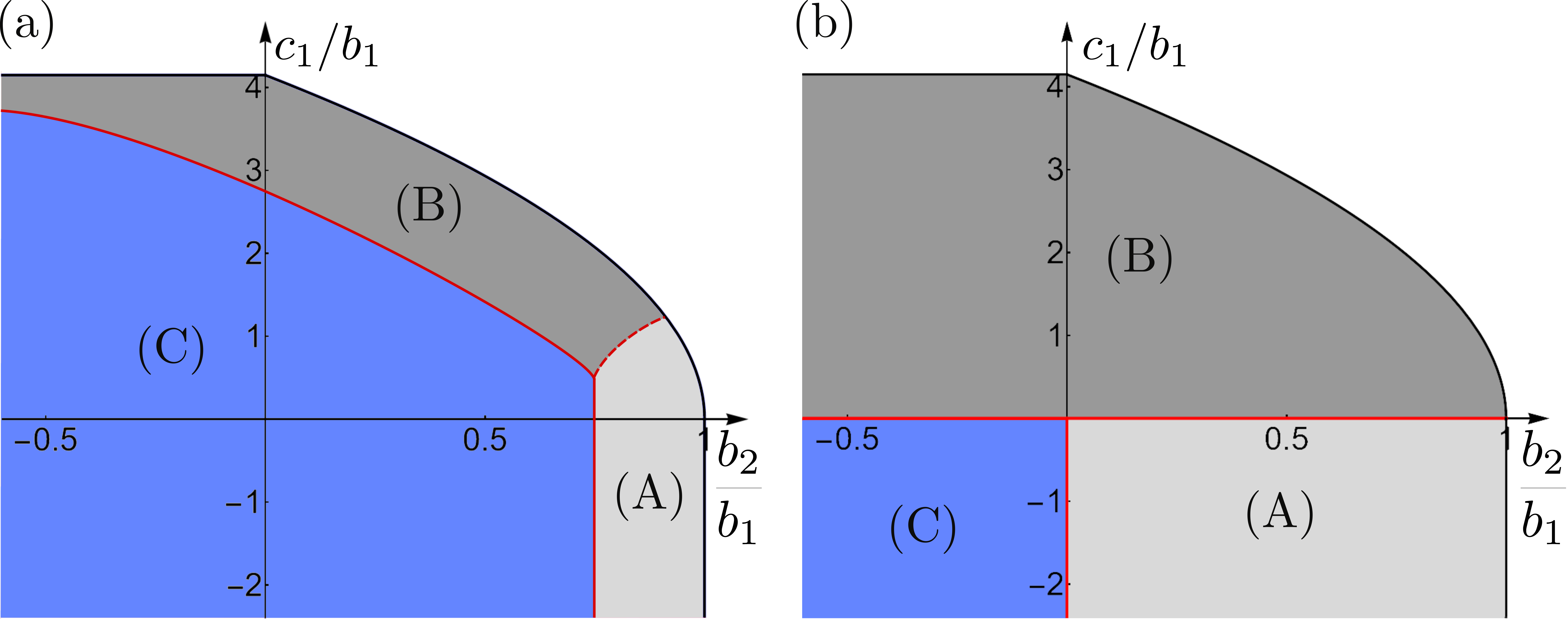}
    \caption{Comparison between (a) the large-$N$ phase diagram in the classical limit and (b) the phase diagram derived by minimizing $V$ in \equref{Potential}. We set $b_1=1$, ${b_3=4.3}$, ${r_d'=0.8}$ and ${r_{N}'=0.7}$. Solid (dashed) red lines refer to first (second) order phase transitions. Lowering temperature in (a) leads to an expansion of the superconducting phases $(A)$ and $(B)$ and the transition between $(A)$ and $(C)$ can become second order as well.}
    \label{fig:PhaseDiagram}
\end{figure}

To allow for the competition of all three phases, we solve the general set of saddle-point equations (\ref{SCE}), which has to be done numerically. A two-dimensional cut through the phase diagram at fixed temperature (fixed $r_d'$, $r_N'$) is shown in \figref{fig:PhaseDiagram}(a). We see that we can indeed obtain all three phases---which of them is favored naturally depends on $b_2/b_1$ and $c_1/b_1$. All transitions shown are first order, except the one between phase $(A)$ and $(B)$, where $\phi_{dN}^0$ sets in continuously. At lower temperatures, the transition between phase $(A)$ and $(C)$ already occurs at lower $b_2/b_1$ and can then become second order too. We have verified that not only phase $(A)$ but also $(B)$ expands in the $b_2/b_1$-$c_1/b_1$ plane, in line with the expectation that lowering temperature increases the tendency of the system to become superconducting. 
For comparison, we show in \figref{fig:PhaseDiagram}(b) the phase diagram obtained simply by minimization of $V$ in \equref{Potential}. Overall, the qualitative orientation of the phases is the same, yet the regions with superconductivity are larger, as expected since fluctuations are neglected. As opposed to the large-$N$ calculation, within this simple, zero-temperature mean-field picture, all phase transitions are first order.

\section{Effective electronic theory}\label{EffElectrTheory}
Having established the phase diagram featuring the two vestigial phases $(A)$ and $(B)$, we will next analyze their electronic properties. Recall that there are four Hubbard-Stratonovich fields, $\psi_d$ (for $\pmb{\bar{d}}\cdot\pmb{d}$), $\psi_N$ (for $\pmb{N}^2$), $\phi_{dd}$ for ($\pmb{d}\cdot\pmb{d}$), and $\phi_{dN}$ for ($\pmb{d}\cdot\pmb{N}$), see \equref{SBHS}, which can be treated on the saddle-point level to leading order within the large-$N$ analysis of \secref{LargeN}. Out of these four saddle-point values, the two associated with $\psi_d$ and $\psi_N$ can be absorbed into a redefinition of the bosonic mass terms $r_{d,N}$ in the respective susceptibility in \equref{BareBosonicAction}. The remaining two, $\phi_{dd}^0$ and $\phi_{dN}^0$, lead to the contribution 
\begin{equation}
    S^0_V=\int_q \left[\phi_{dN}^0\pmb{d}_{q}\cdot\pmb{N}_{-q}+\phi_{d d}^{0} \pmb{d}_{q}\cdot \pmb{d}_{-q}+\text{H.c.}\right]
\end{equation}
in the effective action and characterize the two vestigial phases: 
In phase $(B)$ generally both saddle-point values $\phi_{dd}^0$ and $\phi_{dN}^0$ are expected to be non-zero and complex, whereas in phase $(A)$ only $\phi_{dd}^0\neq 0$. Our primary focus here will be on phase $(A)$, which was not studied before, while we will also present additional results that go beyond the analysis of phase $(B)$ in \refcite{Main_paper}. When discussing phase $(B)$, we will follow \cite{Main_paper} and take the limit of $\phi_{dd}^0=0$ as it does not change any symmetries of the system, offers a more compact discussion of the phase, and can be seen as the limit of large $-b_2$ in \equref{Potential}. 
As such, in our analysis of the system, always only one of the fields $\phi_{dd}^0$ and $\phi_{dN}^0$ will be non-zero, allowing us to pick a gauge in which these fields are real and positive. 

To obtain an effective electronic action, we start from $S = S_e + S_b + S_c + S^0_V$, with the aforementioned saddle-point configurations, and integrate out the bosonic fields. This yields an effective electronic action for phase $(A)$ and $(B)$, which in both cases involves four contributions, $\widetilde{S}^{(A/B)}=S_e+S_1^{(A/B)}+S_2^{(A/B)}+S_{\phi}^{(A/B)}$. Apart from the bare electronic action, $S_e$, there are two particle-number-conserving, and in this sense ``normal'', interactions,
\begin{subequations}\begin{align}
    S_1^{(A/B)}&=\int_{q}p_1^{(A/B)}(q)\left(\pmb{S}_q\pmb{S}_{-q}+\text{H.c.}\right) \label{SpinFlucInt}\\
    S_2^{(A/B)}&=\int_{q}p_{2}^{(A/B)}(q)\left(\pmb{\bar{D}}_q\pmb{D}_q+\text{H.c.}\right), \label{TripletFlucInt}
\end{align}\label{NormalInteractions}\end{subequations}
which capture the contributions from spin fluctuations and superconducting triplet fluctuations, respectively. Here we introduced fermionic bilinears 
\begin{equation}
    \begin{split}
        \pmb{S}_q&=\int_k\bar{c}_{k+q}\pmb{s}\tau_zc_k =\int_{k}\bar{\psi}_{k+q}\pmb{s}\psi_{k},  \\
        \pmb{D}_q&=\int_k c_{k+q}(-is_y\pmb{s})\tau_y c_{-k}=-2i\int_k\bar{\psi}_{k+q}\pmb{s}\gamma_{-}\psi_k,\\
        \pmb{\bar{D}}_q&=\int_k\bar{c}_{k+q}\pmb{s}is_y\tau_y\bar{c}_{-k}=2i\int_k\bar{\psi}_{k+q}\pmb{s}\gamma_{+}\psi_k,\\
    \end{split}
\end{equation}

which we here also already expressed using a Nambu-spinor notation, $\psi_{k}=\left(c_{k,+},is_y\bar{c}_{-k,-} \right)^T$, for later reference. Here $\pmb{\gamma}$ are Pauli matrices acting on Nambu space, and $\gamma_{\pm} = (\gamma_x \pm i \gamma_y)/2$.
As indicated by the superscript $A/B$, the momentum-frequency dependent coupling matrix elements $p_j^{(A/B)}$ depend on which of the two phases we are in. For phase $(A)$, we have
\begin{equation}
    p_{1}^{(A)}(q)=-\frac{\lambda^2}{4\chi^{-1}_{N}(q)}, \,\,\, p_2^{(A)}=-\frac{\lambda^2\chi^{-1}_{d}(q)}{2\left(\chi_{d}^{-2}(q)-4|\phi_{d d}^{0}|^2\right)}, \label{PhaseACouplings}
\end{equation}
meanwhile, for phase $(B)$, it holds
\begin{align}\begin{split}
    p_1^{(B)}(q)&=-\frac{\lambda^2\chi^{-1}_d(q)}{4\left(\chi^{-1}_{d}(q)\chi^{-1}_N(q)-|\phi_{dN}^{0}|^2\right)}, \\ p_2^{(B)}(q)&=-\frac{\lambda^2\chi^{-1}_N(q)}{2\left(\chi^{-1}_d(q)\chi^{-1}_N(q)-|\phi_{dN}^{0}|^2\right)}. \label{PhaseBCouplings}
\end{split}\end{align}
The reason for this difference is rooted in the renormalization of the bosonic propagators coming from the saddle point values $\phi_{dd}^0$ or $\phi_{dN}^0$; indeed, we see that $p_{j}^{(A)} = p_{j}^{(B)}$ in \equsref{PhaseACouplings}{PhaseBCouplings} when setting $\phi_{dd}^0 = \phi_{dN}^0 = 0$.

Finally, for both phases, we also have an ``anomalous'' term, capturing the superconducting nature of the respective saddle point or, more formally, the presence of odd-diagonal-long-range order \cite{penroseBoseEinsteinCondensationLiquid1956, penroseCXXXVIQuantumMechanics1951,yangConceptOffDiagonalLongRange1962,sewellOffdiagonalLongrangeOrder1990}, as demonstrated in \refcite{Main_paper}. Phase (A) is a charge-$4e$ superconductor reflected in its anomalous interaction
\begin{equation}
    S_\phi^{(A)}=\int_{q}\left[ p_{\phi}^{(A)}(q) \pmb{D}_{-q}\pmb{D}_{q}+\text{H.c.}\right] \label{AnomalousInteractionPhaseA}
\end{equation}
consisting of the product of four fermions $cccc$ (and its Hermitian conjugate $c^\dagger c^\dagger c^\dagger c^\dagger$). More intuitively, \equref{AnomalousInteractionPhaseA} describes the fact that the charge-$4e$ superconductivity results from two pairs of electrons, each in a triplet state, forming an overall singlet, as already pointed out above. As opposed to $p_j^{(A)}$ in \equref{PhaseACouplings}, the corresponding coupling function
\begin{equation}
    p_{\phi}^{(A)}(q) = \frac{\lambda^2 \phi_{dd}^{0}}{\left(\chi_{d}^{-2}(q) -4|\phi_{d d}^{0}|^2 \right)}
\end{equation}
is proportional to $\phi_{dd}^{0}$ as required by gauge invariance: under $c_{\vec{k}} \rightarrow e^{i\varphi} c_{\vec{k}}$, we see that $\pmb{D}_{-q}\pmb{D}_{q} \rightarrow e^{4i\varphi }\pmb{D}_{-q}\pmb{D}_{q}$, which is compensated by the factor of $e^{-4i\varphi }$ in $p_{\phi}^{(A)}(q)$ in \equref{AnomalousInteractionPhaseA}, see \tableref{SymmetriesAndReps}. At the same time, this shows that the condensation of $\phi_{dd}$ corresponds to a Higgs mechanism; formally, the difference to the Higgs mechanism in the standard theory of superconductivity is that the residual gauge invariance is $\mathbb{Z}_4$ instead of just $\mathbb{Z}_2$ since the saddle-point action is still invariant under $c \rightarrow i c$ and not just $c \rightarrow - c$.

This is different in phase (B), where the anomalous interaction has the form
\begin{equation}
    S_\phi^{(B)}=\int_{q} \left[ p_{\phi}^{(B)}(q) \pmb{S}_{q}\pmb{D}_q+ \text{H.c.}\right]. \label{AnomalousTermPhaseB}
\end{equation}
Here, $\pmb{S}_{q}\pmb{D}_q$ just picks up a phase of $e^{2i\varphi}$, which is compensated by 
\begin{equation}
    p_{\phi}^{(B)}(q)=\frac{\lambda^2\phi_{dN}^{0}}{2\left(\chi_{d}^{-1}(q)\chi_{N}^{-1}(q) -4|\phi_{d N}^{0}|^2 \right)}
\end{equation}
being proportional to $\phi_{dN}^{0}$, cf.~\tableref{SymmetriesAndReps}. As already discussed at length in \refcite{Main_paper}, this defines a charge-$2e$ superconductor, where a triplet configuration of a particle-hole pair and a triplet Cooper pair form an overall spin-singlet. 

While \equsref{AnomalousInteractionPhaseA}{AnomalousTermPhaseB} are already the analogue of the usual superconducting mean-field order parameter term, they are quartic in the electronic operators. As such, even if we neglected the contribution of the normal interactions (\ref{NormalInteractions}), the fermionic sector of these superconducting states are intrinsically interacting and cannot be solved exactly. This requires using approximate techniques to compute the electronic spectral properties. To this end, we will first study the electronic self energy to leading order in the coupling constant $\lambda^2$, which will then be supplemented by a Hartree-Fock study.

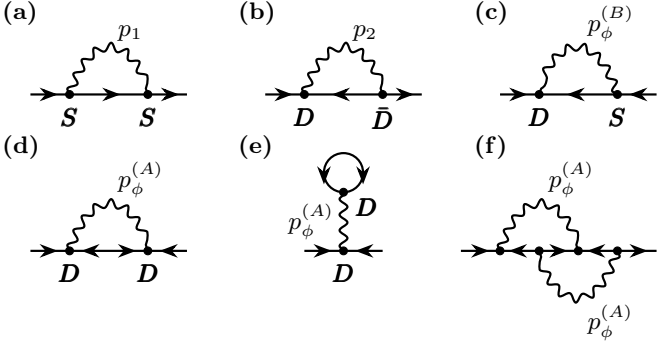
\begin{figure}[t]
    \begin{overpic}[width=0.5\textwidth]{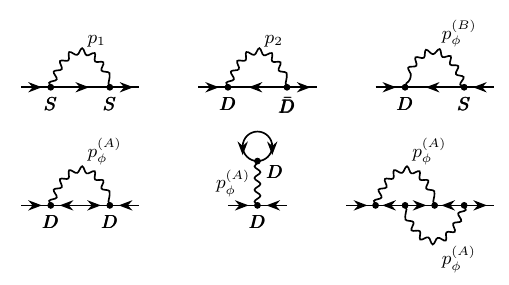}
        \put(0,50){\textbf{(a)}} 
        \put(35,50){\textbf{(b)}} 
        \put(70,50){\textbf{(c)}}
        \put(0,30){\textbf{(d)}} 
        \put(35,30){\textbf{(e)}} 
        \put(70,30){\textbf{(f)}}
    \end{overpic}
     \caption{Diagrams for the perturbative analysis of the electronic spectral function, see text.
     }
     \label{fig:Diagrams}
 \end{figure} 

\subsection{Self-Energy}\label{SelfEnergy}

Starting with the spin-fluctuation contribution (\ref{SpinFlucInt}), spin rotation invariance only allows for the single rainbow diagram shown in \figref{fig:Diagrams}(a) at leading order ($\propto \lambda^2)$; the associated expression for the self-energy reads as  
\begin{equation}
        \Sigma_{1}(k)=6\int_{ \pmb{q}}p_1^{(A,B)}(q)\gamma_zG_{0,k+q}\gamma_z. \label{Sigma1}
\end{equation}
where we used the Nambu representation introduced above with free Green's function given by $G_{0,k}^{-1}=i\omega_n\gamma_0-\epsilon_{\pmb{k}}\gamma_z$. Similarly, the interaction (\ref{TripletFlucInt}) associated with superconducting triplet fluctuations gives rise to a contribution with diagram shown in \figref{fig:Diagrams}(b) and self-energy
\begin{equation}        
    \Sigma_{2}(k)=-24\int_{\pmb{q}}p_2^{(A,B)}(q)\left(\gamma_{-}G_{0,k+q}\gamma_{+}+\gamma_{+}G_{0,k+q}\gamma_{-}\right). \label{Sigma2}
\end{equation}
Note that both $\Sigma_{1}$ and $\Sigma_{2}$ are entirely diagonal in Nambu space, as follows from the matrix structure in \equsref{Sigma1}{Sigma2}, and thus do not induce any anomalous contribution. This, in fact, holds to arbitrary order in perturbation theory for $S_1^{(A/B)}$ and $S_2^{(A/B)}$ by virtue of their charge-conserving form. 

However, for the anomalous interactions $S_\phi^{(A/B)}$ one might expect anomalous self-energies within perturbation theory. This is indeed the case for phase (B), where we obtain the leading order diagram shown in \figref{fig:Diagrams}(c). As the consequences of this term for the spectral functions have already been scrutinized in \refcite{Main_paper}, which was shown to induce a soft gap, we will not study it any further here. Instead, our goal here is to contrast the behavior in phase (A): as a result of the charge-$4e$ nature of the interaction, there is no contribution of $S_\phi^{(A)}$ to order $\lambda^2$. Algebraically, this can be seen by the fact that  
\begin{equation}
        \Sigma_{\phi}(k)=24\int_{\pmb{q}}p_\phi^{(A)}(q)\left(\gamma_{+}G_{0,k+q}\gamma_{+}+\gamma_{-}G_{0,k+q}\gamma_{-}\right)=0
\end{equation}
as a result of the Nambu-space diagonal nature of the bare Green's function $G_0$. We also illustrate this observation diagrammatically in \figref{fig:Diagrams}(d,e), where we show that the contraction of the charge-$4e$ interaction necessarily leads to diagrams involving anomalous internal lines. This shows that at least in the regime where $S_{\phi}^{(A/B)}$ can be treated perturbatively (small $\phi_{dd}^0$ or $\phi_{dN}^0$ and high $T$), there is a qualitative difference between the impact of superconducting correlations in phase $(A)$ and $(B)$---something we will also see in \secref{HartreeFockTheory} on the Hartree-Fock level. We note, of course, that $S_{\phi}^{(A)}$ does lead to contributions in higher orders of perturbation theory, see, e.g., \figref{fig:Diagrams}(f). However, it is easy to see that charge conservation modulo four does not allow for an anomalous self-energy contribution to any order in perturbation theory in phase $(A)$. 

Before we address the physics non-perturbatively below, let us study the impact of the
normal self-energies $\Sigma_{1,2}$ on the electronic spectrum. As we are interested in the behavior at finite temperatures and want to compare the physics with the conceptually related weak-coupling description of the pseudogap in the high-temperature superconductors, we here focus on the ``renormalized classical regime'' \cite{Tremblay2012}. The latter refers to high temperatures and small bosonic mass (large coherence length), where one restricts the Matsubara sums to only the zeroth component. For the cases of an interaction coming from anti-ferromagnetic spin fluctuations, this has been discussed as a way of understanding the pseudogap for finite temperatures and large correlation lengths \cite{PhysRevLett.93.147004}. Also in our case, it will allow for an analytical evaluation of $\Sigma_{1,2}$. 

\begin{figure*}[tb]
    \centering
    \begin{overpic}[trim=6cm 3cm 6cm 3cm, clip,width=1\textwidth]{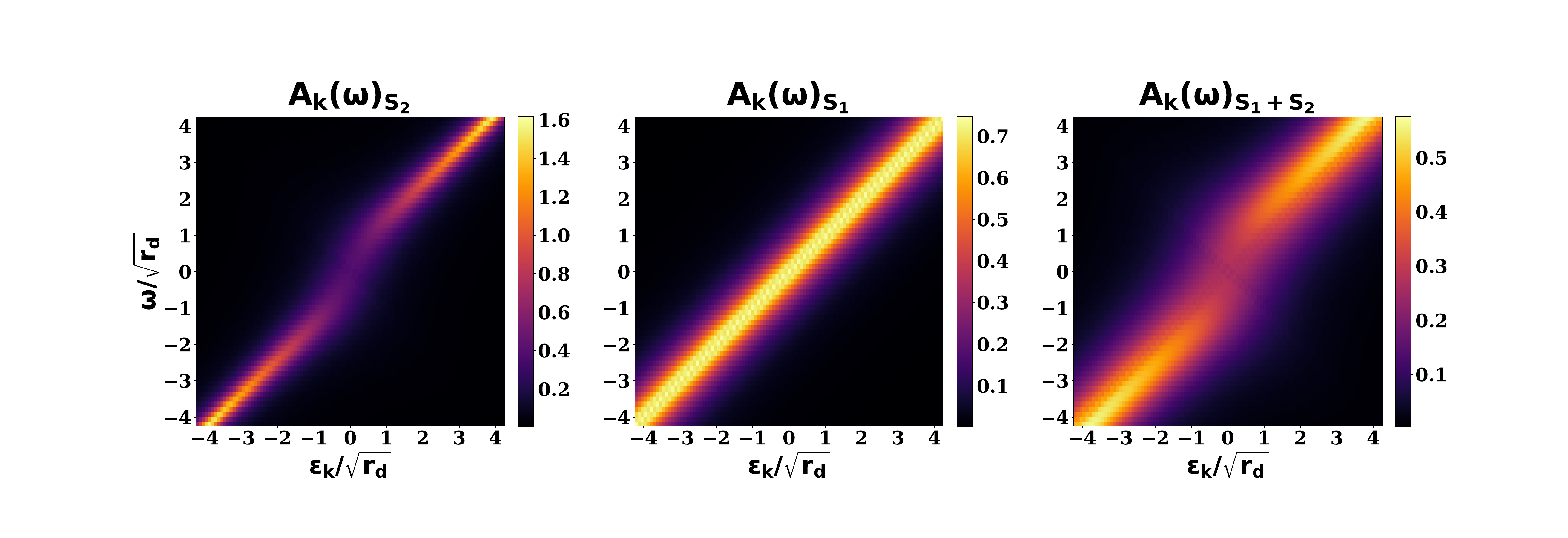}
    \put(5,29){\textbf{(a)}}
    \put(37,29){\textbf{(b)}}
    \put(69,29){\textbf{(c)}}
    \end{overpic}
    \caption{Numerical results for the spectral functions, where the self energy was evaluated including all Matsubara frequencies in the limit of high temperatures and long correlation lengths. \textbf{(a)} Shows the spectral function for the action $S_2$ which describes superconducting triplet fluctuations. One can see an increase in the Fermi velocity as well as life-time broadening near the Fermi surface. \textbf{(b)} Shows the spectral function for the action $S_1$ which describes spin fluctuations. There is no renormalization of the Fermi-velocity, but a life-time broadening for all momenta. \textbf{(c)} Shows the combined effect of $S_1+S_2$ on the spectral function. As expected, there is life-time broadening for all momenta, along with a renormalization of the Fermi-velocity near the Fermi-surface. We chose $\tilde{r}_d/\tilde{r}_N=1$, $v_{d}/v_F=v_{N}/v_F=0.4$, and $\phi^0_{dd}=0$.}
    \label{fig:spectral}
\end{figure*}

To begin with the contribution from superconducting triplet fluctuations, we first note that the coupling $p_3^{(A)}(q)$ in \equref{Sigma2} is replaced by
\begin{equation}
    \tilde{p}_2^{(A)}(\pmb{q})=\left( p_2^{(A)}(q)\right)_{\Omega_n=0}=-\frac{\lambda}{2}\frac{R_d(\vec{q})}{R_d^2(\vec{q})-4(\phi^0_{dd})^2}
\end{equation}
in the renormalized classical regime, where $R_d(\vec{q})=r_d+(v_{d}\pmb{q})^2$; here and in the following, we focus on phase $(A)$ for concreteness, but note that the expressions for phase $(B)$ are closely related and display the same qualitative behavior discussed below.  
After analytic continuation, the particle-particle part of the self-energy reads as
\begin{equation}
    \left(\Sigma_{\pmb{k}}^{(2)}\right)_{11}=-24T\int_{\pmb{q}}\tilde{p}_2^{(A)}(\pmb{q})\frac{1}{\omega+\epsilon_{\pmb{k}+\pmb{q}}+i 0^+}.
\end{equation}
From this, we can readily compute its imaginary part,
\begin{align}
         \mathcal{I}m\left[\Sigma_{\pmb{k}}^{(2)}(\omega)\right]&=24\pi T \int_{\pmb{q}}\tilde{p}_2^{(A)}(\pmb{q})\delta\left(\omega+\epsilon_{\pmb{k}+\pmb{q}}\right) \label{ImPartFirstline}\\
         &=\tilde{g}T \sum_{p=\pm }\frac{1}{\sqrt{\beta_{p}+\left(\omega+ \epsilon_{\pmb{k}}\right)^2}},
\end{align}
where we used $\epsilon_{\pmb{k}+\pmb{q}}\approx \epsilon_{\pmb{k}}+v_Fq_{||}$ with $q_{||}=\pmb{k}\cdot\pmb{q}/|\pmb{k}|$ to arrive at the expression in the second line. Here, we introduced $\tilde{g}=6\lambda \pi v_F/v_d^3$  and $\beta_{\pm}=\left(v_F/v_d\right)^2\left(\tilde{r}_d\pm 2\phi^0_{dd}\right)$. We can see that the imaginary part becomes maximal when $\omega = - \epsilon_{\vec{k}}$, with value becoming large for small $\beta_-$. Note, however, that $\beta_- > 0$ to avoid condensation of the superconducting triplet order parameter, as required by the Mermin Wagner theorem and also reflected by our large-$N$ analysis in \secref{PhaseDiagram} where we found $r_d > 2 \phi^0_{dd}$. 
The real part can be obtained using the Kramers-Kronig relation as
\begin{align}\begin{split}
        &\mathcal{R}e\left[\Sigma_{\pmb{k}}^{(2)}(\omega)\right]=\frac{1}{\pi}\int_{-\infty}^{\infty}\ d\omega' \ \frac{\mathcal{I}m\left[\Sigma_{\pmb{k}}^{(3)}(\omega')\right]}{\omega'-\omega}\\
        &\propto \sum_{p=\pm} \frac{\tilde{g}T}{\sqrt{\beta_{p}+(\omega+ \epsilon_{\pmb{k}})^2}} \ln \left|\frac{\omega+ \epsilon_{\pmb{k}}+\sqrt{\beta_{p}+(\omega+ \epsilon_{\pmb{k}})^2}}{\omega+ \epsilon_{\pmb{k}}-\sqrt{\beta_{p}+(\omega+ \epsilon_{\pmb{k}})^2}}\right|. \label{RealPartofSigma3}
\end{split}\end{align}
We can see that in the immediate vicinity of the maximum of the imaginary part, $|\omega+\epsilon_{\vec{k}}| \ll \beta_p$, the real part of the self-energy is small, since $\mathcal{R}e\left[\Sigma^{(2)}\right] \sim 2 \sum_p \tilde{g}T (\omega+\epsilon_{\vec{k}})/\beta_p$ in this limit from \equref{RealPartofSigma3}. This means that there is no self-energy-induced shift of the band energies right at $\omega = - \epsilon_{\vec{k}}$ and the accompanying maximum of the imaginary part affects the spectral function the most if it coincides with the bare bands, i.e., $\omega = \epsilon_{\vec{k}}$. As such, we have maximum self-energy broadening of the bands in the vicinity of $\epsilon_{\vec{k}}=0$. Intuitively, this related to the fact that the superconducting order parameter (if it was ordered) would affect the electronic spectrum the most right at the Fermi level. This (for $\beta_-\rightarrow 0$ singular) broadening happens at all momenta at the Fermi level since the superconductor is always perfectly ``nested'' on the entire Fermi surface as a result of time-reversal symmetry of the normal-state dispersion. As follows also from $\mathcal{R}e\left[\Sigma^{(2)}\right] \sim 2 \sum_p \tilde{g}T (\omega+\epsilon_{\vec{k}})/\beta_p$, for small but finite $\omega - \epsilon_{\vec{k}}$ the Fermi velocity is enhanced. All of these features are clearly visible in our numerical result for the spectral function shown in \figref{fig:spectral}(a), where we evaluated the self-energy including all Matsubara frequencies in the limit of high temperature and long correlation length (small $\beta_p$).

To contrast this behavior with the well-known pseudogap physics at high temperatures, let us consider anti-ferromagnetic spin-spin fluctuations on the square lattice where the corresponding imaginary part of the self-energy reads in the renormalized classical regime as \cite{Tremblay2012,Pseudogap}
\begin{equation}
    -\mathcal{I}m\left[\Sigma_{\text{AFM}}\right]\propto T\int_{\pmb{q}}\delta\left(\omega -\epsilon_{\pmb{k}+\pmb{q}}\right) \frac{1}{(\pmb{q}-\pmb{Q})^2+\xi^{-2}}. \label{AFMImPart}
\end{equation}
Here $\xi$ is the anti-ferromagnetic correlation length and $\pmb{Q}=(\pi,\pi)^T$ the anti-ferromagnetic momentum transfer. This expression looks very reminiscent of \equref{ImPartFirstline} above, with two main differences: first, there is a relative minus sign of the two terms in the delta-function, which results from the particle-hole and particle-particle nature of $\Sigma_{\text{AFM}}$ and $\Sigma^{(2)}$, respectively. Second, the translational-symmetry breaking of long-range antiferromagnetism leads to the peak of the bosonic propagator in \equref{AFMImPart} being located around $\vec{Q} \neq 0$. This is why, as opposed to the previous case, the maxima of $-\mathcal{I}m\left[\Sigma_{\text{AFM}}\right]$ appear at energies $\omega = \epsilon_{\vec{k} + \vec{Q}}$. For the same reason as above, the impact on the spectral function is the largest if this energy coincides with a bare band-energy, i.e., $\omega = \epsilon_{\vec{k}}$. As such, we obtain significant broadening of the bands whenever $\epsilon_{\vec{k}} = \epsilon_{\vec{k}+\vec{Q}}$. Just as before, this can be viewed as a nesting condition: these are exactly the band energies for which long-range antiferromagnetic order will lead to a gap opening at any finite order parameter strength.

Finally, the self-energy $\Sigma_1$ in \equref{Sigma1}, which comes from ferromagnetic spin fluctuations, can be thought of as the $\vec{Q} \rightarrow 0$ limit of $\Sigma_{\text{AFM}}$ (with, in case of phase $(B)$, slightly more complex bosonic propagator). From the above arguments immediately follows that we expect a finite imaginary part for all energies $\omega = \epsilon_{\vec{k}}$, without any energy/momentum selectivity. This is indeed what can be seen in \figref{fig:spectral}(b), where we show the impact of $\Sigma_1$ on the electronic spectral function. Combining $\Sigma_1$ and $\Sigma_2$, as is relevant for us here, we just obtain a renormalization of the Fermi-velocity around the Fermi-surface as well as lifetime broadening for all momenta which can be seen in Fig. \ref{fig:spectral}(c).

\subsection{Hartree-Fock theory}\label{HartreeFockTheory}
Having established that anomalous contributions in phase $(A)$ can only emerge non-perturbatively, we next address the electronic spectral properties within Hartree-Fock. To this end, we will first cast the fluctuation-induced interactions, see, e.g., \equref{AnomalousInteractionPhaseA}, in the Hamiltonian form, by focusing on the static $i\Omega_n = 0$ contributions. Since it is particularly transparent, we will first further simplify the problem by also setting the momentum transfer of the bosons to zero, $\vec{q}=0$, which can be thought of as taking the limit of a very sharply peaked effective bosonic susceptibility. We will then generalize this by also including $\vec{q} \neq 0$, showing that the $\vec{q}=0$ limit already captures many qualitative features. 

\subsubsection{Minimal Mean-Field Theory for $\pmb{q}=0$} \label{subsec:meanfieldq=0}
In the simplest $\vec{q}=0$ limit, we obtain the effective Hamiltonian contributions for phase $(A)$
\begin{align}
        H_\phi &=  4g_\phi \int_{\pmb{k}_1,\pmb{k}_2} \hspace{-0.3em} (\psi^\dagger_{\pmb{k}_1}\pmb{s}\gamma_+\psi_{\pmb{k}_1})\cdot(\psi^\dagger_{\pmb{k}_2}\pmb{s}\gamma_+\psi_{\pmb{k}_2})+\text{H.c.}, \label{gphiInteraction} \\
        H_2 &= 8g_2\int_{\pmb{k}_1,\pmb{k}_2} (\psi^\dagger_{\pmb{k}_1}\pmb{s}\gamma_+\psi_{\pmb{k}_1})\cdot(\psi^\dagger_{\pmb{k}_2}\pmb{s}\gamma_-\psi_{\pmb{k}_2})
\end{align}
coming from the charge-$4e$ interaction and the particle-number conserving triplet-fluctuation contribution in \equsref{AnomalousInteractionPhaseA}{TripletFlucInt}, respectively.
Just like before with the Grassmann variables, we are using Nambu spinors $\psi_{\pmb{k}}=(c_{\pmb{k},+},is_yc^\dagger_{-\pmb{k},-})^T$ here and further introduced the coupling constants $g_\phi=p_\phi^{(A)}(q=0)$ and $g_2=p_2^{(A)}(q=0)$. Recall that $p_\phi^{(A)} \propto \phi_{dd}^0$ and hence transforms non-trivially under a U(1) gauge transformation. However, as pointed out above, we here choose a gauge where $\phi_{dd}^0$ and, thus, $g_\phi$ are real and positive. We note that there is an ambiguity in the choice of our Hamiltonian as one moves from the Grassmann algebra to the fermionic algebra. We pick our Hamiltonian such that the mean-field calculations qualitatively agree with the results obtained from the spectral analysis and that the self-consistency equations obtained from the Hamiltonian coincide with the saddle point equations after performing a Hubbrd-Stratonovich transformation. As we have seen in \secref{SelfEnergy} that ferromagnetic spin fluctuations have only a minor qualitative effect on the electronic spectrum, we neglect it here for simplicity.

Together with the free contribution, $H_0=\int_{\pmb{k}} \psi_{\pmb{k}}^{\dagger}\epsilon_{\pmb{k}} \gamma_z \psi_{\pmb{k}}$, the total interacting Hamiltonian is given by $H_{\text{(A)}} = H_0 + H_{\phi} + H_2$. We next perform a mean-field decoupling of the interaction terms, only allowing for contractions that retain spin-rotation invariance as required by the Mermin-Wagner theorem. Defining the $2 \times 2$ (in the particle-hole space) correlation matrix $\left(\underline{\pmb{C}}_{\pmb{k}}\right)_{ij}=\braket{(\psi_{\pmb{k}}^{\dagger})_j (\psi_{\pmb{k}})_i}$, the resulting mean-field Hamiltonian can be written as
\begin{equation}
    \begin{split}
        H_{\text{MF}}&=H_0-24\int_{\pmb{k}} \psi^\dagger_{\pmb{k}}\Bigl(g_\phi\left[\gamma_{+}\cdot\underline{\pmb{C}}_{\pmb{k}}\cdot\gamma_{+}+\text{H.c.}\right]\\
        &\quad +g_2\left(\left[\gamma_{-}\cdot\underline{\pmb{C}}_{\pmb{k}}\cdot\gamma_{+}+\text{H.c.}\right]-\gamma_{+}\cdot \gamma_{-}\right)\Bigr)\psi_{\pmb{k}}\\
    \end{split}
    \end{equation}
We now parameterize the different components of the correlator as
\begin{equation}
    \underline{\pmb{C}}_{\pmb{k}}=\frac{1}{2}\gamma_0+C^z_{\pmb{k}}\gamma_z-\Im[C_{\pmb{k}}]\gamma_y+\Re[C_{\pmb{k}}]\gamma_x, \label{CParameterization}
\end{equation}
where $C^z_{\pmb{k}} \in \mathbb{R}$, $C_{\pmb{k}} \in \mathbb{C}$.
After inserting this ansatz into $H_{\text{MF}}$ we obtain 
\begin{equation}
    H_{\text{MF}}=\int_{\pmb{k}} \psi^\dagger_{\pmb{k}}\begin{pmatrix}
            \tilde{\epsilon}_{\pmb{k}}&-\Delta^*_{\vec{k}}\\
            -\Delta_{\vec{k}}&-\tilde{\epsilon}_{\pmb{k}}\\
        \end{pmatrix}\psi_{\pmb{k}}\\
\end{equation}
where $\tilde{\epsilon}_{\pmb{k}}=\epsilon_{\pmb{k}} + \delta \epsilon_{\pmb{k}}+12g_2$, with $\delta \epsilon_{\pmb{k}}=24g_2C^z_{\pmb{k}}$, is the renormalized dispersion and $\Delta_{\vec{k}} = 24 g_\phi C_{\vec{k}}$ the superconducting, charge-$2e$ order parameter. An additional constant diagonal term $12g_2\gamma_z$ can be absorbed into the chemical potential and therefore will not be mentioned in the following considerations. 

To derive the self-consistency relations, we recompute the correlation function from $H_{\text{MF}}$ at temperature $\beta^{-1}$,
\begin{equation}
    \begin{split}           \left(\underline{\pmb{C}}_{\pmb{k}}\right)_{ij}&=\mathbb{I}_{ij}-\frac{\Tr\left[(\psi_{\pmb{k}})_i(\psi^{\dagger}_{\pmb{k}})_je^{-\beta H_{\text{MF}}}\right]}{\Tr\left[e^{-\beta H_{\text{MF}}}\right]}\\
    \end{split}
\end{equation}
to obtain
\begin{equation}
    \underline{\pmb{C}}_{\pmb{k}}=\frac{1}{2}\gamma_0+\frac{1}{4}t(\beta E_{\pmb{k}}/2) \beta \begin{pmatrix}
    -\tilde{\epsilon}_{\pmb{k}}&\Delta_{\vec{k}}^*\\
    \Delta_{\vec{k}}&\tilde{\epsilon}_{\pmb{k}}\\
    \end{pmatrix}
\end{equation}
where $t(x) = \tanh(x)/x$ and $E_{\pmb{k}}=\sqrt{\tilde{\epsilon}_{\pmb{k}}^2+|\Delta_{\vec{k}}|^2}$. Combining this with \equref{CParameterization}, we obtain the self-consistency relations given by
\begin{subequations}
\label{SelfConsistencyEquations1}\begin{align}
    \Delta_{\pmb{k}} &= 6 \beta g_\phi t(\beta E_{\pmb{k}}/2) \Delta_{\pmb{k}}^*, \label{DeltaSCequation} \\
    \delta \epsilon_{\pmb{k}} &=-6 \beta g_2 t(\beta E_{\pmb{k}}/2) (\epsilon_{\pmb{k}} + \delta \epsilon_{\pmb{k}}). \label{SelfConsistencyEquationDEps}
\end{align}\end{subequations}
Note that, as opposed to the usual superconducting self-consistency equations, we have $\Delta$ on one and $\Delta^*$ on the other side of \equref{DeltaSCequation}. This is in fact required by gauge invariance as the coupling constant $g_{\phi} \propto \phi_{dd}^0$ carries charge-$4e$, see \equref{gphiInteraction}. As mentioned before, we here choose a gauge where $g_{\phi} > 0$ leading to $\Delta_{\pmb{k}} \in \mathbb{R}$. 
\begin{figure*}[tb]
    \centering
    \begin{overpic}[width=1\linewidth]{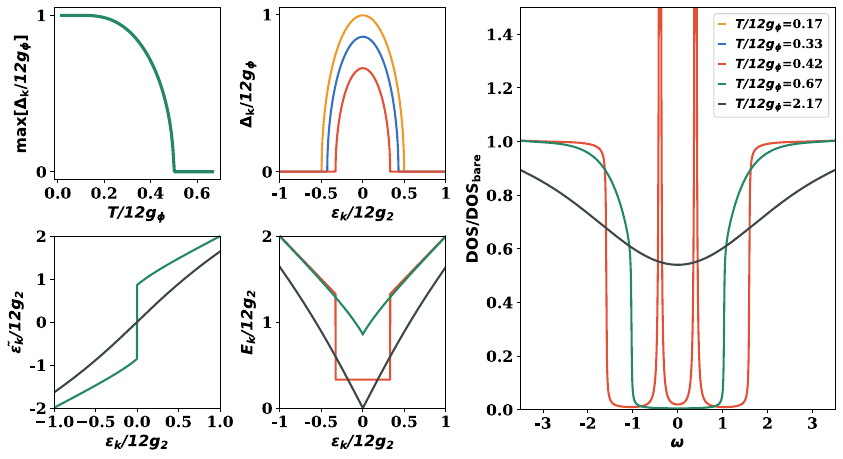}
    \put(2,55){\textbf{\rmfamily(a)}}
    \put(29,55){ \textbf{\rmfamily(b)}}
    \put(2,28){ \textbf{\rmfamily(c)}}
    \put(29,28){ \textbf{\rmfamily(d)}}
    \put(57,55){ \textbf{\rmfamily(e)}}
    \end{overpic}
    \caption{Results for phase $(A)$ with $|g_2/g_{\phi}|=2$, i.e., $T^* > T_c$. In \textbf{(a)} we show the behavior of the maximum value of the superconducting order parameter $\Delta_{\pmb{k}}$ with respect to the temperature. The solutions for the self-consistent Hartree-Fock calculations are shown in \textbf{(b)-(d)}. The superconducting order parameter \textbf{(b)} is non zero for larger regions around the Fermi surface for lower temperatures. The renormalized dispersion \textbf{(c)} has a discontinuity for temperatures below $T^{*}$, which can also be seen in the dispersion of the excitations \textbf{(d)} where a gap is found for $T<T^{*}$. For temperatures below $T_c$ one can see that the dispersion becomes completely flat in the region where $\Delta_{\pmb{k}}\neq 0$.  \textbf{(e)} The DOS for temperatures in the three regimes $T>T^*$ (black), $T_c<T<T^*$ (green) and $T<T_c$ (red). For $T<T_c$ one obtains $\delta$-like coherence peaks (regularized when including finite $\vec{q}$, see \figref{fig:selfconsistencyq}) inside the larger region of completely suppressed DOS. Due to the gap in the dispersion, the DOS still vanishes in an entire energy range around the Fermi level for $T_c<T<T^*$; when $T>T^*$, one only finds a partial suppression  of the DOS at low energies, leading to a $V$-shaped behavior.}
    \label{fig:self_consistencyregime1}
\end{figure*}
\begin{figure*}[tb]
    \centering
    \begin{overpic}[width=0.8181\textwidth]{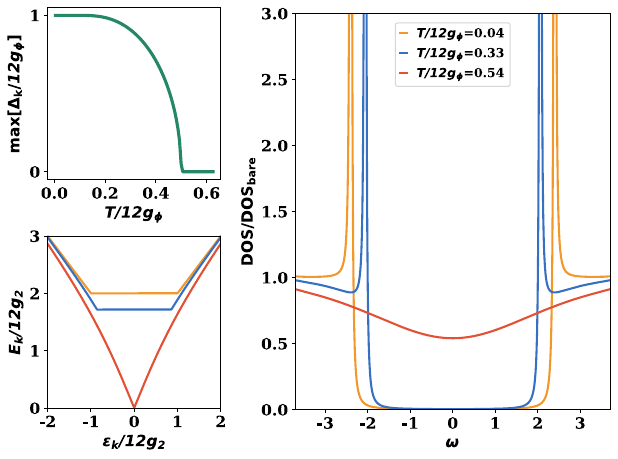}
    \put(2,75){\textbf{(a)}}
    \put(2,38){\textbf{(b)}}
    \put(41,75){\textbf{(c)}}
    \end{overpic}
    \caption{Results for phase $(A)$ for $|g_2/g_{\phi}|=0.5$, meaning $T_c>T^*$. \textbf{(a)} shows the maximum of $\Delta_{\vec{k}}$ as a function of $T$, revealing that $\Delta_{\vec{k}}$ only becomes non-zero below $T_c$. The dispersion of the excitations \textbf{(b)} does not have a gap for $T>T_c$ (red), which only emerges below $T_c$ (blue and orange). The corresponding DOS \textbf{(c)} shows $\delta$-like coherence peaks for $T<T_c$, which are regularized by finite $\vec{q}$, cf.~\figref{fig:selfconsistencyq}. Due to the renormalization of the Fermi velocity near the Fermi surface, the low-energy DOS is partially suppressed already for $T>T_c$.}
    \label{fig:self_consistencyregime2}
\end{figure*}

If we assume that $\Delta_{\pmb{k}} \neq 0$, at a given $\vec{k}$, \equref{DeltaSCequation} can be rewritten as
\begin{equation}
    \frac{1}{6\beta g_\phi} = t(\beta E_{\pmb{k}}/2). \label{DeltaNonzeroCond}
\end{equation}
As $t(x) \leq 1$, this equation can only be satisfied if $6\beta g_\phi \geq 1$, i.e., at sufficiently strong coupling and low temperature. As such, finite $\Delta_{\vec{k}}$ can only develop below the critical temperature $T_{c} = 6g_{\phi}$, as can also be seen in our numerical solution shown in \figref{fig:self_consistencyregime1}(a). Note that this is consistent with our perturbative analysis in \secref{SelfEnergy}, where no anomalous two-particle correlations were induced in phase $(A)$ to arbitrary order. We can now see that---at least within the mean-field approximation---non-perturbative effects become relevant at (and below) $T_{c}$. What is more, as $t(x)$ is a monotonically decreasing function of $|x|$, we can see that $\Delta_{\vec{k}} \neq 0$ first develops at the Fermi level and, at fixed $\beta g_\phi$, is only finite around it, see \figref{fig:self_consistencyregime1}(b).

Apart from $\Delta_{\vec{k}}$, there is also a renormalization of the normal-state bands, governed by the second self-consistency equation (\ref{SelfConsistencyEquationDEps}) and driven by the interaction $g_2$ (like in our perturbative approach). We can clearly see that $\delta \epsilon_{\vec{k}} \neq 0$ is generically expected and not just below a certain temperature, including above $T_{c}$ where $\Delta_{\vec{k}}=0$. In fact, in this regime, we can further simplify \equref{SelfConsistencyEquationDEps} to 
\begin{equation}
    \delta \epsilon_{\pmb{k}} =-12g_2 \tanh(\beta \tilde{\epsilon}_{\pmb{k}}/2). \label{NormalStateRenorm}
\end{equation}
Recalling that $g_2=4p_2^{(A)}(q=0) < 0$, see \equref{PhaseACouplings}, we see that $\sign \, \delta \epsilon_{\pmb{k}} = \sign \, \tilde{\epsilon}_{\pmb{k}}$ and, as such, expect that it enhances the Fermi velocity, in line with the perturbative result in \figref{fig:spectral}(a). In fact, expanding \equref{NormalStateRenorm} to leading order in $\tilde{\epsilon}_{\pmb{k}}$, i.e., close to the renormalized Fermi surface, one can derive the explicit expression
\begin{equation}
    \tilde{\epsilon}_{\pmb{k}} \sim \left(1 + \frac{6 \beta |g_2|}{1 - 6 \beta |g_2|} \right) \epsilon_{\vec{k}}, \quad \epsilon_{\vec{k}} \rightarrow 0, \label{BandRenormalization}
\end{equation}
for the renormalized dispersion (valid for $6 \beta |g_2| < 1)$. This renormalization is clearly visible in the high-temperature regime of our numerical solution of \equref{SelfConsistencyEquationDEps}, see black curve in shown in \figref{fig:self_consistencyregime1}(c). As signaled by the divergent denominator in \equref{BandRenormalization}, at lower temperature, we obtain an additional, non-perturbative solution. To see this, we note that \equref{NormalStateRenorm} can be rewritten as 
\begin{equation}
    \frac{1}{6 \beta |g_2|} = t(\beta \delta \epsilon_{\vec{k}}/2) \label{EquationForTstar}
\end{equation}
for $\epsilon_{\vec{k}} = 0$, which has the same structure as \equref{DeltaNonzeroCond}. We see that a non-zero $\delta \epsilon_{\vec{k}}$ and hence a gap is obtained at the Fermi level if $\beta |g_2| \geq 1/6$ or, equivalently, for temperatures below $T^* = 6 |g_2|$, in line with our numerics in \figref{fig:self_consistencyregime1}(c). Despite the similarities of \equsref{DeltaNonzeroCond}{EquationForTstar}, there is a crucial physics distinction: while the onset of finite $\max_{\vec{k}}|\Delta_{\vec{k}}|$ actually corresponds to a phase transition (from charge-$4e$ to charge-$2e$ superconductivity), the emergence of $\delta \epsilon_{\vec{k}}|_{\epsilon_{\vec{k}}=0} \neq 0$ does not; in this sense, the mathematical similarities are just a consequence of our simplified theoretical description that only takes into account $\vec{q}=0$. 

Nonetheless, the presence of two temperature scales---$T_c$ and $T^*$---does lead to complex physics. The ratio of these temperatures is determined by the interaction constants, $T^{*}/T_c=|g_2/g_\phi|$, and although our large-$N$ analysis in~\secref{PhaseDiagram} and Mermin-Wagner theorem suggest that $r_d > 2 \phi^0_{dd}$ guaranteeing $|g_{2} / g_{\phi}| > 1$, we find it instructive to investigate the spectrum in both regimes, $|g_2/g_\phi|>1$ and $|g_2/g_\phi|<1$. While \figref{fig:self_consistencyregime1} refers to the former regime, we contrast it with the latter in \figref{fig:self_consistencyregime2}. We can see by comparing \figref{fig:self_consistencyregime1}(d) and \figref{fig:self_consistencyregime2}(b) that the temperature dependence of the electronic excitation spectrum $E_{\pmb{k}}$, which depends on both $\tilde{\epsilon}_{\pmb{k}}$ and $\Delta_{\vec{k}}$, differs: in the regime of $|g_2/g_\phi|>1$, one can observe a gap in the electronic excitation spectrum induced by the $H_2$-interaction before the onset of the superconducting order parameter $\Delta_{\pmb{k}}$ when $T_c<T<T^*$. This feature does not appear in the other regime, as the onset of the superconducting order parameter means the excitation spectrum is dictated by the self consistency (\ref{DeltaNonzeroCond}) for all $T<T_c$ and not by \equref{EquationForTstar}. A striking feature in both regimes is that $E_{\pmb{k}}$ becomes perfectly flat for energies $|\tilde{\epsilon}_{\vec{k}}|$ smaller than $|\Delta_{\vec{k}}|$. This can be straightforwardly understood from~\equref{DeltaNonzeroCond} by noting that it explicitly depends only on the $E_{\vec{k}}$, meaning its solution is independent of $\vec{k}$ for all $\vec{k}$ with $\Delta_{\vec{k}} \neq 0$. We note, however, that this feature is a result of the simple approximation of only including $\vec{q}=0$, as we will see below. Another interesting feature is that for $T < T_{c}$ the interplay of $H_{\phi}$ and $H_{2}$ interactions leads to the discontinuity of $E_{\vec{k}}$ at the bare energy $\epsilon_{\vec{k}}$ where $\Delta_{\vec{k}}$ vanishes as illustrated by \figref{fig:self_consistencyregime1}(d) (red curve) and \figref{fig:self_consistencyregime2}(b) (blue and orange curves).

In \figref{fig:self_consistencyregime1}(e) and \figref{fig:self_consistencyregime2}(c), we plot the resulting DOS for varying temperatures. Due to the perfectly flat Bogoliubov bands in the current $\vec{q}=0$ approximation, one obtains $\delta$-like coherence peaks in both regimes. These peaks are located inside the region of completely suppressed DOS. This is in stark contrast to the BCS-like spectrum where coherence peaks coincide with the edges of the vanishing DOS region. We again note that this feature can be attributed to the interplay of the interactions since the BCS-like result can be recovered by setting $g_{2}$ to zero. Due to the band renormalization increasing the Fermi velocity near the Fermi surface, which occurs in both regimes, there is a partial, $V$-shaped suppression of the DOS around the Fermi surface for higher temperatures ($T > T_c$), see, e.g., black curve in \figref{fig:self_consistencyregime1}(e). The appearance of a gapped excitation spectrum for $T_c<T<T^*$ in the regime $|g_2/g_{\phi}|>1$, causes a complete suppression of the DOS around the Fermi surface, see green curve in \figref{fig:self_consistencyregime1}(e).

\begin{figure*}[tb]
    \centering

    \begin{overpic}[width=0.9\textwidth]{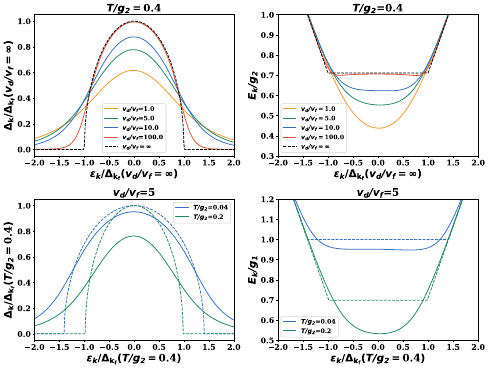}
    \put(2,73){\textbf{(a)}} 
    \put(49,73){\textbf{(b)}} 
    \put(2,35){\textbf{(c)}} 
    \put(49,35){\textbf{(d)}} 
    \end{overpic}
    \caption{Results for phase $(A)$, including finite-$\vec{q}$ contributions. In all the figures the dashed lines always show the solutions for $\pmb{q}=0$. In the figures \textbf{(a)} and \textbf{(b)} the temperature is fixed, where as in the figures \textbf{(c)} and \textbf{(d)} the relation between the Fermi-velocity and $v_d/v_f$ is fixed. Also the solutions for finite $\pmb{q}$ washes out the abrupt change in behavior of the solution between the regime where the superconducting gap is zero and non-zero in the case of $\pmb{q}=0$. \textbf{(a)}  The superconducting gap for different $v_d/v_f$ and for finite $\pmb{q}$ converges towards the solution for $\pmb{q}=0$ in the limit of $v_d/v_f\rightarrow \infty$. \textbf{(b)} The dispersion of the Bogoliobuv excitations for different $v_d/v_f$ and for finite $\pmb{q}$ converges towards the solution for $\pmb{q}=0$ for $v_d/v_f\rightarrow \infty$. For $\pmb{q}=0$ we see that the dispersion is completely flat in the region where the superconducting gap is non-zero. \textbf{(c)} The superconducting gap $\Delta_{\pmb{k}}$ for different temperatures, where we see that increasing the temperature increases the washing out effect. \textbf{(d)} The dispersion for the Bogoliobov excitations for different temperatures.}
    \label{fig:selfconsistencyq}
\end{figure*}

Finally, we will contrast the $\vec{q}=0$ mean-field theory for phase $(A)$ with that of phase $(B)$. The latter was already discussed in \refcite{Main_paper}, so we will directly state the corresponding self-consistency equations here,
\begin{subequations}
\begin{align}
    \Delta_{\pmb{k}} &= \beta\tilde{g}_{\phi} t(\beta E_{\pmb{k}}/2) \tilde{\epsilon}_{\pmb{k}}, \\
    \tilde{\epsilon}_{\pmb{k}} &= \epsilon_{\pmb{k}} +\beta \tilde{g}_{\phi} t(\beta E_{\pmb{k}}/2) \Delta_{\pmb{k}}.
\end{align}\label{0qphaseBMF}\end{subequations}
Here $\tilde{g}_\phi$ is the effective Hamiltonian coupling constant associated with the interaction $S_\phi^{(B)}$ in \equref{AnomalousTermPhaseB}, and all other interactions are neglected for simplicity. We see a rather different structure compared to \equref{SelfConsistencyEquationDEps} since now both self-consistency equations (\ref{0qphaseBMF}) involve $\Delta_{\vec{k}}$ but each only on one side. This structure is dictated again by gauge invariance by noting that the interaction with coupling constant $\tilde{g}_\phi$ carries charge $2e$.

One immediate consequence of this structure is that we obtain finite $\Delta_{\vec{k}}$ at any temperature, as opposed to what we found above for phase $(A)$---again in line with our perturbative discussion in \secref{SelfEnergy}. As analyzed in detail in \refcite{Main_paper} and we will revisit below when generalizing these equations to finite $\vec{q}$, the high-temperature regime is indeed perturbative, with $\Delta_{\vec{k}}$ vanishing at the Fermi level; this leads to a soft gap. At lower temperature, \equref{0qphaseBMF} has additional non-perturbative solutions with a hard gap. However, the fact that this appears as a non-perturbative solution at a specific temperature is---similar to $\delta \epsilon_{\vec{k}}|_{\epsilon_{\vec{k}}=0}$ above---just a consequence of our fine-tuned theoretical description and not a signal of an additional phase transition, as we will see more explicitly in \secref{Finite2PhaseB}.

\subsubsection{Finite $\pmb{q}$ for Phase $(A)$}\label{FiniteqPhaseA}

After gaining intuition within our analytically transparent $\vec{q}=0$ mean-field theory, we go beyond this approximation and include finite momenta. As before, we start with phase $(A)$. To keep the discussion compact, we focus on the effect of the anomalous interaction in \equref{AnomalousInteractionPhaseA} to study the onset of charge-$2e$ superconducting correlations and neglect the band renormalization emerging from the particle-number conserving interaction in Eq.~\eqref{TripletFlucInt}, as our numerical results show that the phase-like transition to a gapped excitation spectrum for $T_c<T<T^*$ does not appear when considering finite momentum. As expected, this is a feature that arises from taking the approximation of setting $\pmb{q}=0$.

Denoting the static limit $p_{\phi}^{(A)}(q)|_{i\Omega_n\rightarrow 0}$ of the coupling function by $\bar{p}_{\phi}^{(A)}(\pmb{q})$, the resulting mean-field Hamiltonian becomes
\begin{equation}
        H_{\text{MF}}^{(A)}=H_0-24\int_{\pmb{q},\pmb{k}}\hspace{-0.1em}\bar{p}_{\phi}^{(A)}(\pmb{q})\psi^\dagger_{\pmb{k}}\Bigl(\gamma_{+}\cdot\underline{\pmb{C}}_{\pmb{k}+\pmb{q}}\cdot\gamma_{+} + \text{H.c.} \Bigr)\psi_{\pmb{k}}.\\
\end{equation}
Since we set $p_2^{(A)}(q)=0$, there is no band renormalization and therefore $\bar{\epsilon}_{\pmb{k}}=\epsilon_{\pmb{k}}$. We hence have only one self-consistency equation to solve,
\begin{equation}
    \D=6\int_{\pmb{q}} \beta \bar{p}_{\phi}^{(A)}(\pmb{q}) t(\beta E_{\pmb{k}}/2) \Delta^*_{\pmb{k}+\pmb{q}}, \label{FiniteqDeltaEquation}
\end{equation}
with $E_{\pmb{k}}=\sqrt{\epsilon_{\pmb{k}}^2+\left(\sum_{\pmb{q}}24p_\phi^{(A)}(\pmb{q})\Delta_{\pmb{k}+\pmb{q}}\right)^2}$.
Equation~(\ref{FiniteqDeltaEquation}) is the generalization of \equref{DeltaSCequation} beyond $\bar{p}_{\phi}^{(A)}(\pmb{q}) = \delta(\vec{q}) g_\phi$. 
To be able to compare our results with the $\vec{q}=0$ limit, we will normalize $\bar{p}_{\phi}^{(A)}$ in the following by setting 
\begin{equation}
    \sum_{\pmb{q}}\bar{p}_{\phi}^{(A)}(\pmb{q})=g_\phi.
\end{equation}
As before, we will choose a gauge with $\bar{p}_{\phi}^{(A)}(\pmb{q}) \in \mathbb{R}^+$, which implies $\Delta_{\vec{k}} \in \mathbb{R}$. Equation (\ref{FiniteqDeltaEquation}) then assumes the usual form of a mean-field gap equation. As such, we know that there is still a critical temperature $T_c$ above (below) which $\Delta_{\vec{k}} = 0$ ($\Delta_{\vec{k}} \neq 0$). At this temperature, the system transitions from a charge-$4e$ to a charge-$2e$ superconductor. As can be seen in \figref{fig:selfconsistencyq}(a,c), the presence of finite momentum transfer washes out the existence of two regions in momentum space with $\Delta_{\vec{k}} = 0$ and $\Delta_{\vec{k}} \neq 0$. We can see that increasing the bosonic velocity $v_d$ decreases this broadening effect, which is expected as the susceptibility will get increasingly peaked. In fact, one can see in \figref{fig:selfconsistencyq}(a) that we recover the $\vec{q}=0$ limit for $v_d/v_f \rightarrow \infty$ (where $v_f$ is the Fermi velocity). Finally, the Bogoliubov excitation spectrum shown in \figref{fig:selfconsistencyq}(b,d) reveals that the flat region (and associated $\delta$-like peak in the DOS) we found in the $\vec{q}=0$ limit is regularized.\par

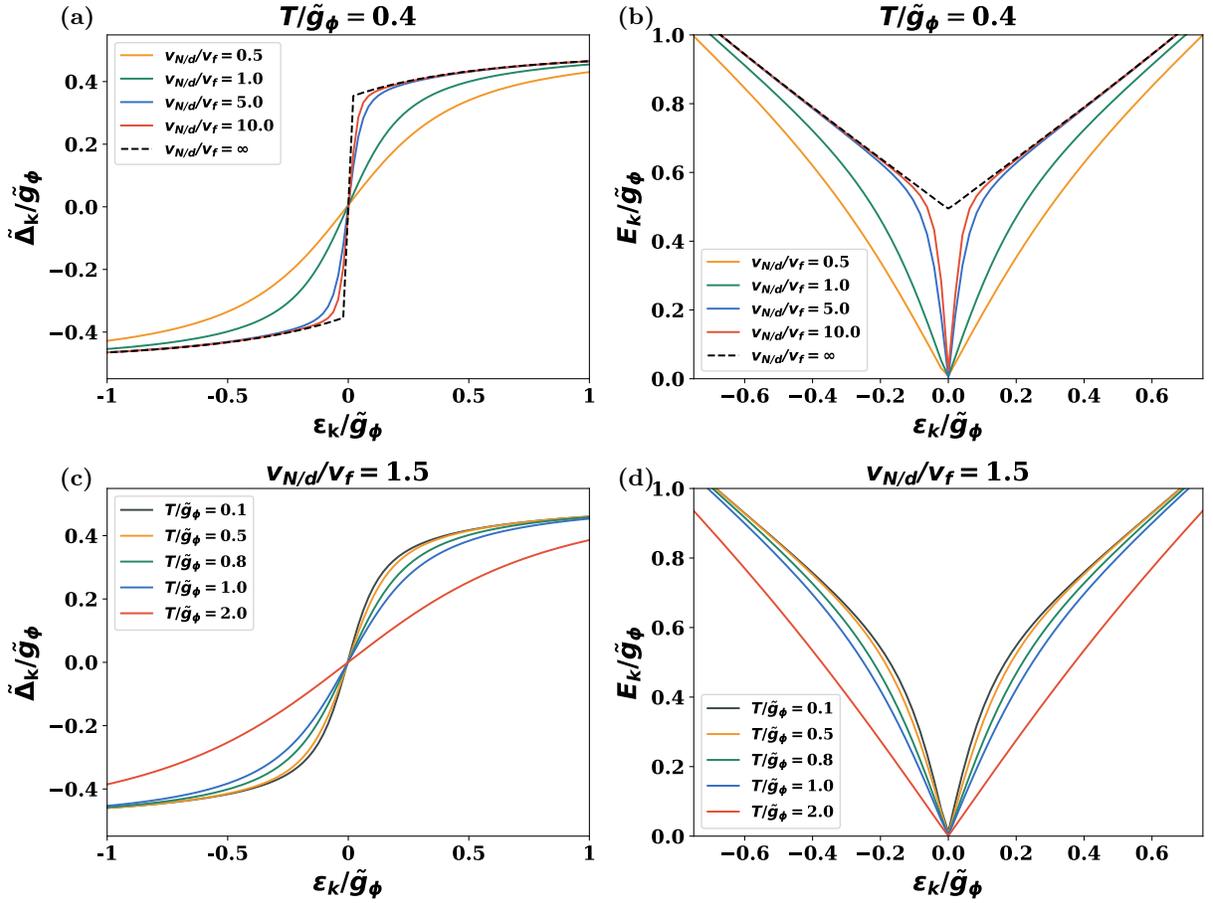
\begin{figure*}[tb]
    \centering
    \begin{overpic}[width=0.9\textwidth]{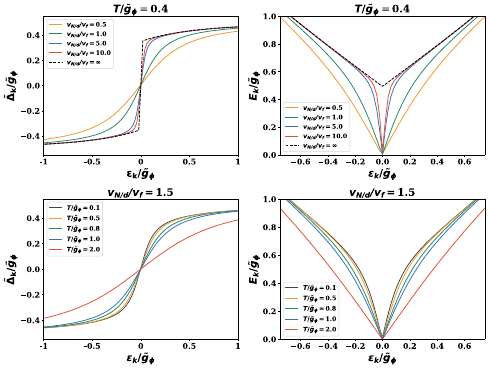}
    \put(5,73){\textbf{(a)}} 
    \put(51,73){\textbf{(b)}} 
    \put(5,35){\textbf{(c)}} 
    \put(51,35){\textbf{(d)}} 
    \end{overpic}
    \caption{Results for phase $(B)$, including finite-$\vec{q}$ contributions. In the figures \textbf{(a)} and \textbf{(b)} the temperature is fixed, where as in the figures \textbf{(c)} and \textbf{(d)} the relation between the Fermi-velocity and $v_{N/d}/v_f$ is fixed. \textbf{(a)}  The superconducting gap for different $v_{N/d}/v_f$ and finite $\pmb{q}$ converges towards the solution for $\pmb{q}=0$ in the limit of $v_{N/d}/v_f\rightarrow \infty$. Finite $\pmb{q}$ also washes out the hard features found in the $\pmb{q}=0$ approximation. \textbf{(b)} The dispersion of the bogoliobuv excitations for different $v_{N/d}/v_f$ and for finite $\pmb{q}$ converges towards the solution for $\pmb{q}=0$ for $v_{N/d}/v_f\rightarrow \infty$. \textbf{(c)} The superconducting gap $\Delta_{\pmb{k}}$ for different temperatures, where we see that increasing the temperature increases the washing out effect. \textbf{(d)} The dispersion for the bogoliobov excitations for different temperatures.}
    \label{fig:selfconsistency_SD}
\end{figure*}

\subsubsection{Finite $\pmb{q}$ for Phase $(B)$}\label{Finite2PhaseB}
Finally, we will consider phase $(B)$, which was only discussed in the $\pmb{q}=0$ approximation in \refcite{Main_paper}. Focusing, again, only on the anomalous interaction $S_\phi^{(B)}$ in \equref{AnomalousTermPhaseB} and using $\bar{p}_{\phi}^{(B)}(\pmb{q})$ to denote the static limit of $p_{\phi}^{(B)}(q)|_{i\Omega_n\rightarrow 0}$, the self-consistency equations read as
\begin{subequations}
\begin{align}
    \Delta_{\vec{k}} &= \frac{1}{4} \int_{\pmb{q}} \beta \bar{p}_{\phi}^{(B)}(\pmb{q}) t(\beta E_{\pmb{k}+\vec{q}}/2) \tilde{\epsilon}_{\vec{k}+\vec{q}}, \\
    \tilde{\epsilon}_{\vec{k}} &= \epsilon_{\vec{k}} + \frac{1}{4} \int_{\pmb{q}} \beta \bar{p}_{\phi}^{(B)}(\pmb{q}) t(\beta E_{\pmb{k}+\vec{q}}/2) \Delta_{\vec{k}+\vec{q}},
\end{align}\label{FiniteqPhaseB}\end{subequations}
with the modified Bogoliubov dispersion $E_{\pmb{k}}=\sqrt{\tilde{\epsilon}_{\pmb{k}}+\Delta_{\pmb{k}}}$. As required, these equations reduce to \equref{0qphaseBMF} in the limit $\bar{p}_{\phi}^{(B)}(\pmb{q}) = \delta(\vec{q}) \tilde{g}_\phi$.

Figure~\ref{fig:selfconsistency_SD} illustrates the numerical solutions of \equref{FiniteqPhaseB}. We see that the main effect of finite $\pmb{q}$ is to wash out the hard features that appear for $\pmb{q}=0$. For increasing $v_{N/d}/v_F$, the solution of the superconducting gap and the dispersion of the Bogoliubov excitations converge towards the solution obtained in the $\pmb{q}=0$ approximation, represented as dashed lines, as shown in \figref{fig:selfconsistency_SD}(a) and \figref{fig:selfconsistency_SD}(b). The behavior of the superconducting gap and of the Bogoliubov excitations with respect to the temperature is shown in \figref{fig:selfconsistency_SD}(c) and \figref{fig:selfconsistency_SD}(d), respectively.
As anticipated above, we can see that the non-perturbative solution forming a hard gap at low temperatures disappears.

\section{Conclusions}\label{Summary}
In this work, we studied the finite-temperature phase diagram and the superconducting spectral properties of a two-dimensional system which has strong tendencies towards both triplet pairing (order parameter $\vec{d}$) and spin magnetism ($\vec{N}$). 
The phase diagram, obtained within a large-$N$ approach, is shown in \figref{fig:PhaseDiagram}(a) and features two superconducting phases: phase $(A)$---a charge-$4e$ state characterized by the condensation of $\vec{d}\cdot\vec{d}$, while $\braket{\vec{d}}=\braket{\vec{N}}=\braket{\vec{d}\cdot\vec{N}}=0$---and phase $(B)$, where $\braket{\vec{d}\cdot\vec{N}}\neq 0$, which is best thought of as the condensation of three electrons and a hole, forming a spin-singlet, charge-$2e$ boson. The comparison with the phase diagram in \figref{fig:PhaseDiagram}(b), which is obtained simply by minimization of the interaction potential in \equref{Potential} and can thus be seen as a simple form of mean-field analysis at zero temperature, shows that the location of these two states in the phase diagram is energetically natural. Meanwhile, we also see that the finite-temperature fluctuations included in \figref{fig:PhaseDiagram}(a) lead to an expansion of the non-superconducting region $(C)$, as expected. The phase transition between the two superconductors $(A)$ and $(B)$ is found to be second order, while both the thermal phase transitions as well as the transition between $(C)$ and $(A)$ can be first or second order depending on parameters. 

When describing the electronic properties of the two superconducting states, a crucial complication, compared to the standard theory of superconductivity, is that even the analogue of the celebrated mean-field approach---here replaced by the large-$N$ theory and treating the Hubbard-Stratonovich fields $\phi_{dd}$ and $\phi_{dN}$ at the saddle-point level---leads to an interacting fermionic theory. This involves the particle-number conserving interactions in the spin-spin and triplet-pairing-triplet-pairing channel in \equref{NormalInteractions}, as well as an anomalous interaction; for instance, for phase $(A)$, the latter is the charge-$4e$ contribution in \equref{AnomalousInteractionPhaseA}. The reason for the non-quadratic nature of the saddle-point theories is that, for both phases, the elementary Cooper pairs are replaced by bound states of four particles. 

This is why approximate methods are needed to compute the electronic spectral function. We started with a perturbative computation of the self-energy. For phase $(A)$, there is no anomalous self-energy contribution, to any order in perturbation theory, as a result of particle-number conservation modulo four. This is to be contrasted with phase $(B)$, where such an anomalous contribution arises already in first order perturbation theory since the associated interaction, see \equref{AnomalousTermPhaseB}, conserves the particle number only modulo two. We also compare the self-energy contributions of the two normal interactions (\ref{NormalInteractions}) with the well-known case of anti-ferromagnetic spin fluctuations \cite{Tremblay2012} with momentum transfer $\vec{Q}=(\pi,\pi)^T$ on the square lattice. While the latter gives rise to a significant suppression of the spectral weights at the ``hot spots’’, where $\epsilon_{\vec{k}} = \epsilon_{\vec{k}+\vec{Q}}$, the superconducting triplet fluctuations lead to a suppression at the Fermi surface, $\epsilon_{\vec{k}}=0$, where superconductivity is ``nested’’. Finally, fluctuations of $\vec{N}$ lead to equal broadening of the entire band and, hence, no interesting momentum dependence, which follows from the lack of momentum transfer, $\vec{Q}=0$, associated with $\vec{N}$.

We then complement this perturbative approach with a mean-field treatment in order to capture non-perturbative effects in a compact way. For instance, the associated self-consistency equations for phase $(A)$ in the simplest limit only with vanishing momentum transfer ($\vec{q}=0$) are shown in \equref{SelfConsistencyEquations1}. Notably, the equation for the superconducting charge-$2e$ order parameter involves $\Delta_{\vec{k}}$ on one and $\Delta_{\vec{k}}^*$ on the other side, as dictated by the charge-$4e$ nature of the interaction in \equref{AnomalousInteractionPhaseA} or \equref{gphiInteraction} and thus of the associated coupling constant $g_\phi$. 
Importantly, it only has a solution with $\Delta_{\vec{k}} \neq 0$ below a certain temperature. This is in agreement with the absence of an anomalous self-energy within perturbation theory and shows that the charge-$4e$ state is stable and only develops charge-$2e$ singlet pairing at sufficiently low temperature and/or large $g_\phi$. The second equation in \equref{SelfConsistencyEquations1} describes a band renormalization that is present at any temperature and leads to a soft suppression of the DOS with decreasing temperature before a hard gap sets in. We showed that these features remain qualitatively valid when finite $\vec{q}$ are taken into account; the same applies for phase $(B)$, where the key spectral properties described in \refcite{Main_paper} are still present, albeit ``washed out’’ in momentum space.  

Overall, our work shows that systems with strong tendencies towards triplet superconductivity and magnetism, such as graphene-based moiré superlattices, are very promising platforms for exotic vestigial phases with interesting spectral properties. Natural next steps are to generalize this theory to momentum-dependent superconducting order parameters and to analyze the electronic spectra with other numerical techniques such Monte-Carlo methods or dynamical mean-field theory. \par

\begin{acknowledgments}
D.S.~and M.S.S.~acknowledge funding by the European Union (ERC-2021-STG, Project 101040651---SuperCorr). Views and opinions expressed are however those of the authors only and do not necessarily reflect those of the European Union or the European Research Council Executive Agency. Neither the European Union nor the granting authority can be held responsible for them. 
\end{acknowledgments}

\bibliography{apssamp}
\end{document}